# Atomic-Scale Investigation of an Asymmetric $SrTiO_3$ Grain Boundary

Janina Malin Rybak[1], Jonas Arlt[1], Qian Ma[1], Carmen Fuchs[2], Baptiste Gault[3,4], Timo Jacob[2], Christian Jooss[1], Tobias Meyer[1], Cynthia A. Volkert[1]

*1 Institute of Materials Physics, University of Göttingen, Friedrich-Hund-Platz 1, 37077 Göttingen*
*2 Institute of Electrochemistry, Ulm University, Albert-Einstein-Allee 47, 89081 Ulm*
*3 Max Planck Institute for Sustainable Materials, Max-Planck-Straße 1, 40237 Düsseldorf*
*4 Department of Materials, Imperial College London, Kensington, London, SW7 2AZ, United Kingdom*

**Abstract:** Grain boundaries (GBs) in oxide perovskites significantly influence their functional properties. This study examines the atomic-scale structure and composition of a faceted asymmetric grain boundary in strontium titanate ($SrTiO_3$) using scanning transmission electron microscopy (STEM), atom probe tomography (APT), and density functional theory (DFT). STEM and APT reveal an atomically sharp boundary with asymmetric and symmetric facets, marked by strong Sr depletion over a width of less than 1 nm. STEM-EELS shows Ti concentration variations of up to 20% between facets, while APT constrains this variation to less than 10%. DFT+*U* calculations of a symmetric Σ5 facet confirm that Sr depletion minimizes boundary energy while maintaining Ti content. The variation in Ti suggests cation mobility that enables local energy minimization. Differences in facet surface energies likely drive Ti redistribution, offering strategies for GB structure control. This combined experimental-theoretical approach provides key insights into the structure and energy landscape of GBs in $SrTiO_3$, aiding in the prediction of their impact on ionic and electronic transport.

## 1. Introduction

Grain boundaries (GBs) in perovskite oxides play a critical role in determining the functional properties of these materials, with $SrTiO_3$ (STO) serving as a model system for investigating these effects. The properties of STO are highly sensitive to stoichiometry and defect structures, particularly at grain boundaries, where local structural and chemical variations can significantly impact electronic and ionic transport, catalytic activity, and dielectric behavior. While GBs are often associated with detrimental effects, they also provide opportunities for engineering desired functionalities, making a comprehensive atomic-scale understanding essential for optimizing STO-based materials for applications such as capacitors[1] and resistive switching devices.[2] Furthermore, understanding the GB structure and impact on ion migration in STO as a perovskite model system allows insights into the complex behavior of solid oxide fuel cells.[3]

A key factor influencing GB behavior in STO is related to the mixed ionic and covalent bonding, which leads to complex local distortions of oxygen octahedra and a rich variability of point defects and non-stoichiometry.[4,5] These atomic-scale variations modify the local energy landscape, altering bonding environments, valence states, and electrostatic fields.[6] The presence of space charge layers at GBs can create large internal fields, strongly affecting the transport of charged species both along and across the boundary.[7] In addition, where GBs intersect the material surface, they can introduce variations in catalytic activity, further emphasizing the need for detailed structural and chemical characterization at the atomic level.[8]

Transmission electron microscopy (TEM) and density functional theory (DFT) have played a central role in advancing our understanding of grain boundaries in STO and other perovskite oxides. TEM-based investigations have provided near-atomic-resolution insights into the structure, defect distribution, and electrostatic characteristics of GBs, while DFT has complemented these studies by elucidating the local energy landscape and bonding environments. Early TEM studies relied on high-resolution TEM (HRTEM) and negative-Cs imaging (NCSI) to infer oxygen vacancy concentrations and ordering in STO bulk and twin boundaries.[9,10] More recently, annular bright-field scanning TEM (ABF-STEM) and integrated differential phase contrast (iDPC) STEM have improved the detection of light elements such as oxygen, allowing direct visualization of octahedral rotations at oxide interfaces.[11–13] Additionally, techniques such as electron holography (EH) and momentum-resolved STEM (MRSTEM) have been employed to map electrostatic potentials across GBs at the atomic and nanoscale.[14,15]

Despite these advances, several limitations persist in TEM-based GB studies. Most notably, the inherent 2D nature of regular (S)TEM imaging means that full 3D atomic-scale reconstructions remain challenging. Optical sectioning with nanometer depth resolution is hindered by microscope hardware limitations,[16,17] while electron channeling effects complicate direct interpretation of images, particularly in systems with compositional or structural heterogeneity.[18] Recent advances in electron ptychography, a powerful 4D-STEM technique, have demonstrated the potential to overcome these limitations by enabling 3D atomic-scale imaging through numerical reconstruction of diffraction data.[19] However, these methods require specialized instrumentation and extensive computational resources and have so far been applied primarily in simulation studies of oxide perovskite interfaces.[20] Additionally, spectroscopic techniques such as electron energy loss spectroscopy (EELS) and energy-dispersive X-ray spectroscopy (EDX) have been used to probe local stoichiometry at GBs.[15,21] However, their quantitative interpretation remains challenging due to multiple scattering, channeling effects, and a lack of depth resolution, particularly in facetted GBs containing both light and heavy elements.[18,22]

The limitations of TEM-based methods in chemical sensitivity and three-dimensional resolution motivate the use of atom probe tomography (APT), which provides near-atomic-scale 3D compositional mapping. APT is a powerful technique for analyzing the chemical composition of features such as grain boundaries and the spatial distribution of trace and light elements, making it an ideal complement to TEM-based methods.[23,24] However, conventional APT analysis of perovskites has been challenging due to the brittle nature of these materials, which tend to fracture under the high electric fields required for measurement.[25,26]

Until recently, reported APT measurement success rates for perovskites were lower than 10% to 20%. Even when successful, the datasets obtained were typically too small to provide meaningful insights. This issue has been observed in studies of highly Nb-doped bulk STO,[27,28] STO thin films,[29,30] as well as bulk BTO specimens[31] and nanoparticles.[26] However, recent advancements have demonstrated that applying 10–20 nm conformal metal coatings to atom probe tips significantly reduces fracture incidence in brittle materials.[31,32] This improvement is likely due to the coating's ability to reduce electrostatic fields and mechanical stresses within the specimen during field evaporation.

Very recently, this approach has enabled successful APT measurements of undoped and Nb-doped STO, confirming compositional homogeneity in the bulk material down to the nanoscale, with a precision of ±1% for Sr, Ti and O.[33] These advancements mark a significant step forward in the atomic-scale characterization of perovskite oxides, overcoming previous barriers to high-resolution chemical mapping in these complex materials.

Theoretical modeling provides a crucial counterpart to experimental studies, allowing us to extend and generalize the understanding of GB structures beyond individual experimental observations. By modeling the local energy landscape, charge distribution, and defect interactions at GBs, density functional theory (DFT) can help explain observed structural distortions and electronic property variations. When DFT predictions match experimental results, it provides a certain degree of confidence that these models can even be used to predict the behavior of other GBs in STO and similar materials. For example, prior DFT studies have successfully reproduced electrostatic potential variations and defect segregation patterns observed in STEM and EH studies.[34] Furthermore, DFT enables the calculation of migration barriers for ionic species, helping to quantify how GBs influence ionic transport, defect energetics and space-charge effects.[35–37] This predictive capability is essential for optimizing STO-based materials for applications in resistive switching, catalysis, and thermoelectrics, where the behavior of GBs is a key determinant of performance.[38]

In this study, we investigate the atomic-scale structure of a faceted asymmetric $\Sigma5$ tilt grain boundary in STO using a combination of advanced characterization and computational techniques. Scanning transmission electron microscopy (STEM) is employed to resolve atomic arrangements and provide chemical insights, while atom probe tomography (APT) offers a three-dimensional chemical composition analysis. Density functional theory (DFT) calculations complement these experimental approaches by revealing the energy landscape governing atomic positions and migration barriers. By integrating these techniques, we establish a detailed, atomic scale understanding of the asymmetric GB in STO, contributing to the broader goal of controlling and engineering GB properties in perovskite oxides.

## 2. Materials and Methods

### 2.1. Sample Material

All studies have been carried out on a commercially manufactured $SrTiO_3$ (STO) diffusion-bonded bicrystal containing an asymmetric ≈ 36.9° [100] tilt grain boundary (CRYSTAL GmbH, Berlin, Germany). The undoped crystals were grown by the Verneuil method and then diffusion bonded to produce a mostly pore-free, atomically sharp boundary.[6] Diffusion bonding of STO is typically performed at temperatures in excess of 1400 °C and over several hours,[39] providing sufficient conditions for diffusion of the intrinsic species over distances larger than investigated here is this study.

The bicrystal features a single [100] tilt grain boundary, which is perpendicular to the (100) sample surface and has a tilt angle of approximately 36.9°. The boundary is flat with symmetric $\Sigma 5$ and asymmetric facets of a few nm length. It contains a few voids, around 10 nm in size. The median grain boundary plane is tilted ~ 8.9° and ~ 28° with respect to the [010] directions of the two crystals, i.e. the median plane is tilted ~ 9.6° off a nominally symmetric (013) plane (see Figure S4).

### 2.2. Atom Probe Tomography

APT specimens were prepared from the near surface regions of the bulk bicrystal in a Thermo Fisher Scientific Helios G4 dual beam SEM/FIB instrument. The grain boundary of the bicrystal was identified in SEM images based on voids left by the diffusion bonding process and marked with *e*-beam deposited Pt. A micrometer size block of STO material was extracted from the surface and attached to a W support stub using FIB-aided Pt deposition (see Figure 1). The STO block was then sharpened into a cone-shaped tip with an apex radius of 15–30 nm using lateral FIB cuts, guided by reference measurements to position the grain boundary along the tip's central axis.[40] After final cleanup with a 5 kV, 15 pA ion beam from above, the sharpened specimens were coated with a thin (10 to 30 nm) Cr layer via redeposition from milling a Cr target inside the FIB, near the apex of the APT specimen.[32,33] This approach is essential for collecting enough data during APT measurement to obtain high-resolution chemical information from STO.

The APT experiments were performed using a custom-built laser-assisted wide angle atom probe (for details see Maier et al.[41]) with a 133 mm straight flight path and a conventional large-diameter electrode. The laser was operated at a wavelength of $\lambda$ = 355 nm with a pulse length of 15 ps, a focal spot size of approximately $\omega_{1/2}$ = 150 μm, and repetition rate of 100 kHz. The base temperature was set to 120 K and the laser pulse energy to 0.11 µJ. The detection rate window was maintained at 0.15–0.3 ions per 100 pulses. Data reconstruction and analysis were performed using Scito 2.3.6 (Inspico),[42] with further analysis conducted using the open-source software 3depict 0.0.23[43] and custom python scripts. The reconstruction was performed using the point-projection protocol by Geiser et al.[44] with shank angle and initial radius values obtained from TEM and refined by aligning the STO core dimensions in the reconstructed APT dataset with TEM images.

## 2.3. Transmission Electron Microscopy

TEM plan-view specimens of thicknesses between 40 and 50 nm were prepared with the FIB following the method described by Jublot and Texier[45] and using a final Ga ion acceleration voltage of 2 kV during sample thinning. Subsequent conventional TEM and STEM measurements were performed in a FEI Titan 80-300 E-TEM operated at 300 kV. A beam current of approximately 40 pA and a convergence semi-angle of 10 mrad were used during STEM. High-angle annular dark-field (HAADF) images were recorded with a collection semi-angle of 47-100 mrad. EELS data was acquired with a Gatan GIF Quantum 965 ER using a dispersion of 1 eV/channel and a collection semi-angle of 37 mrad. Resulting core-loss signals were extracted after power law background subtraction.

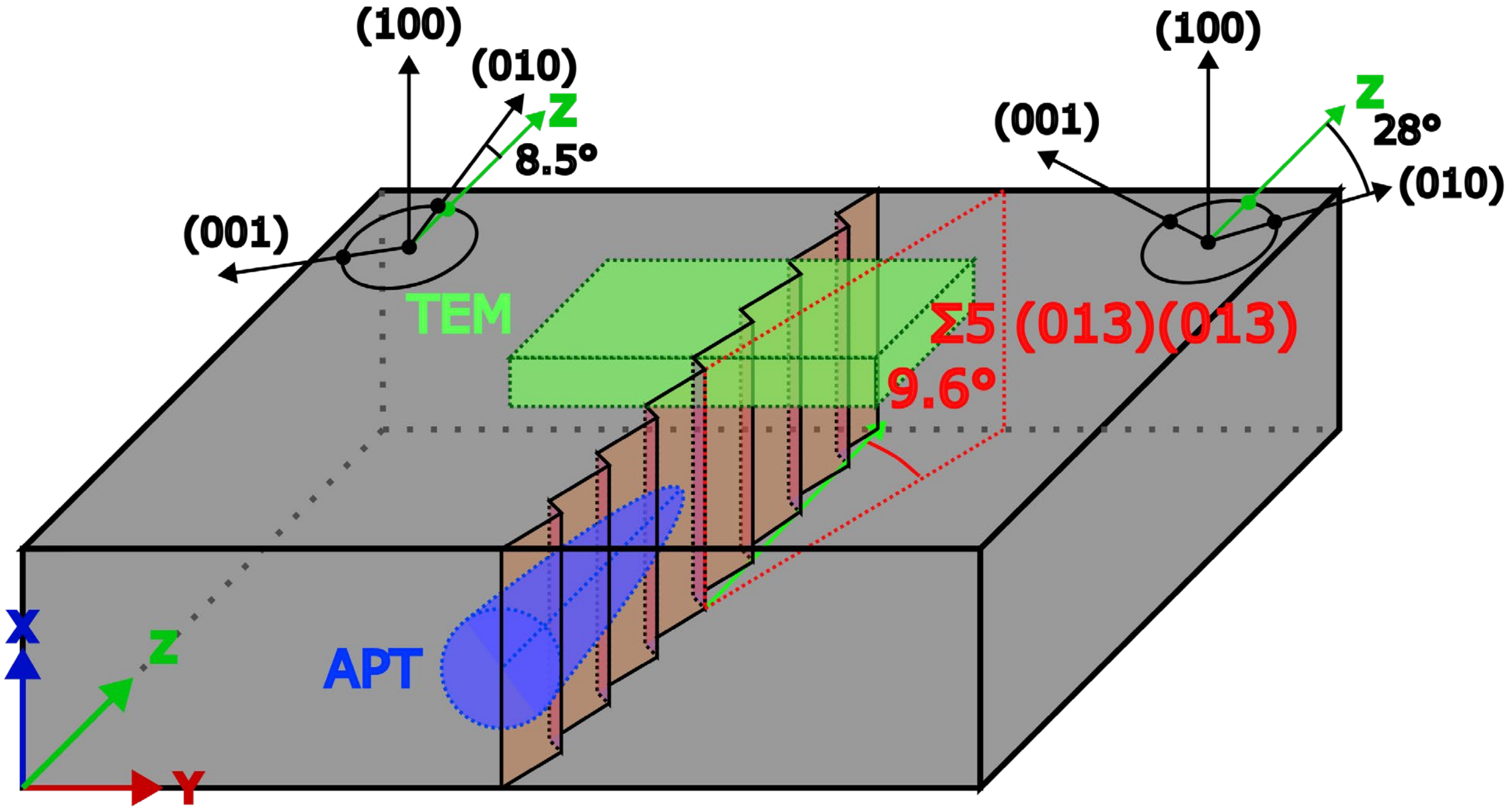

***Figure 1:*** *Schematic of the sample geometry showing the relation between the crystal directions of the grains, the grain boundary, and the coordinate system of the TEM and APT specimens and for the DFT calculations.*

## 2.4. Density Functional Theory

Periodic density functional theory (DFT) simulations of several STO surfaces and grain boundaries were performed using the plane-wave-based Vienna *Ab Initio* Simulation Package (VASP)[46,47]. Core electrons were represented through the projector augmented-wave (PAW) method developed by Blöchl[48] as implemented in VASP.[49] In this approach, the semi-core *s*- and *p*-states were treated as part of the valence electron configuration. To account for exchange and correlation effects, the Generalized Gradient Approximation (GGA) based on the functional suggested by Perdew, Burke, and Ernzerhof (PBE)[50] was employed. Spin-polarization was included in all calculations and the plane-wave energy cutoff was set to 500 eV. For Brillouin zone sampling, a Monkhorst–Pack[51] ***k***-point mesh with a density of 0.20 $Å^{-1}$ was applied. A Gaussian smearing of $\sigma = 0.01$ eV was employed, and the self-consistent field (SCF) cycle was deemed converged when the total energy difference was less than $10^{-4}$ eV. For both bulk and surface calculations, the geometry was considered converged when the forces on all atoms were below 1.0 meV·$Å^{-1}$, which was increased to 10 meV·$Å^{-1}$ for grain boundary calculations. In order to ensure that an adequate description of the electronic properties of the system is provided, the coulombic Hubbard *U* correction[52] was employed on the *d*-shell of Ti, as well as on the *p*-states of oxygen with values of $U_d$ = 8.0 eV and $U_p$ = 6.0 eV, respectively. As this work employs the rotationally invariant formulation by Liechtenstein *et al*.[53] the exchange interactions *J* were included with values of $J_d$ = 0.6 eV and $J_p$ = 0.1 eV.

The optimized STO bulk structure was used to generate symmetric surface slabs representing the (100), (110), (013) and (021) surfaces of strontium titanate. A 14 Å vacuum layer was introduced on both sides of the slab to eliminate interactions between the periodic slabs. Depending on the surface orientation, different morphologies are possible: the (100) surface can exhibit either $TiO_2$ termination (100):$TiO_2$ or SrO termination (100):SrO, while the (110) surface is stable with either $SrTiO_2$ (110):$SrTiO_2$ or O termination (110):O. The (013) and (021) surfaces adopt a stoichiometric configuration. To model possible non-stoichiometries at the surfaces, supercells containing between 104 and 124 atoms were constructed for each surface, incorporating varying amounts of Ti, Sr, and O atoms.

A symmetric Σ5[100](013) GB was generated from the cubic perovskite unit cell, as illustrated schematically in Figure 2. The (013) surface, which had already been generated and optimized according to the method described above, was employed to make up the left grain of the boundary. Tilting the left grain by 18.45° followed by a mirroring step generated the right grain (see Figure 2). In order to optimize the GB structure, the following procedure was employed: Initially, the two grains were separated by a distance of two STO lattice constants, *i.e.* that differences in evaporation field at the macroscopic level of the boundary plane remain below the sensitivity of our measurements. However, local variations at individual facets cannot be ruled out. 2*a* = 7.92 Å in *y*-direction. The distance was then gradually reduced in steps of 0.05*a* while generating structures. A merging threshold of 1.5 Å was set to ensure atomic spacing remained within tolerance, meaning atoms closer than this threshold were merged by removing one of the competing atoms. Next, adjustments in *z*-direction were made by varying the separation in steps of 0.1*a*, applying the same distance tolerance. To reduce computational cost, the *y*-dimension (normal to the GB) was optimized first, followed by the *z*-dimension (along the GB and perpendicular to the main tilt rotation axis), minimizing the number of structural variations and ensuring efficient use of resources. After relaxation, the minimum energy structure was used for further studies. Convergence assessments were conducted to ascertain the influence of adjacent grain boundaries on the supercell size. The interaction was regarded as negligible when the coordination of the innermost atom in each grain approximates its bulk phase counterpart.

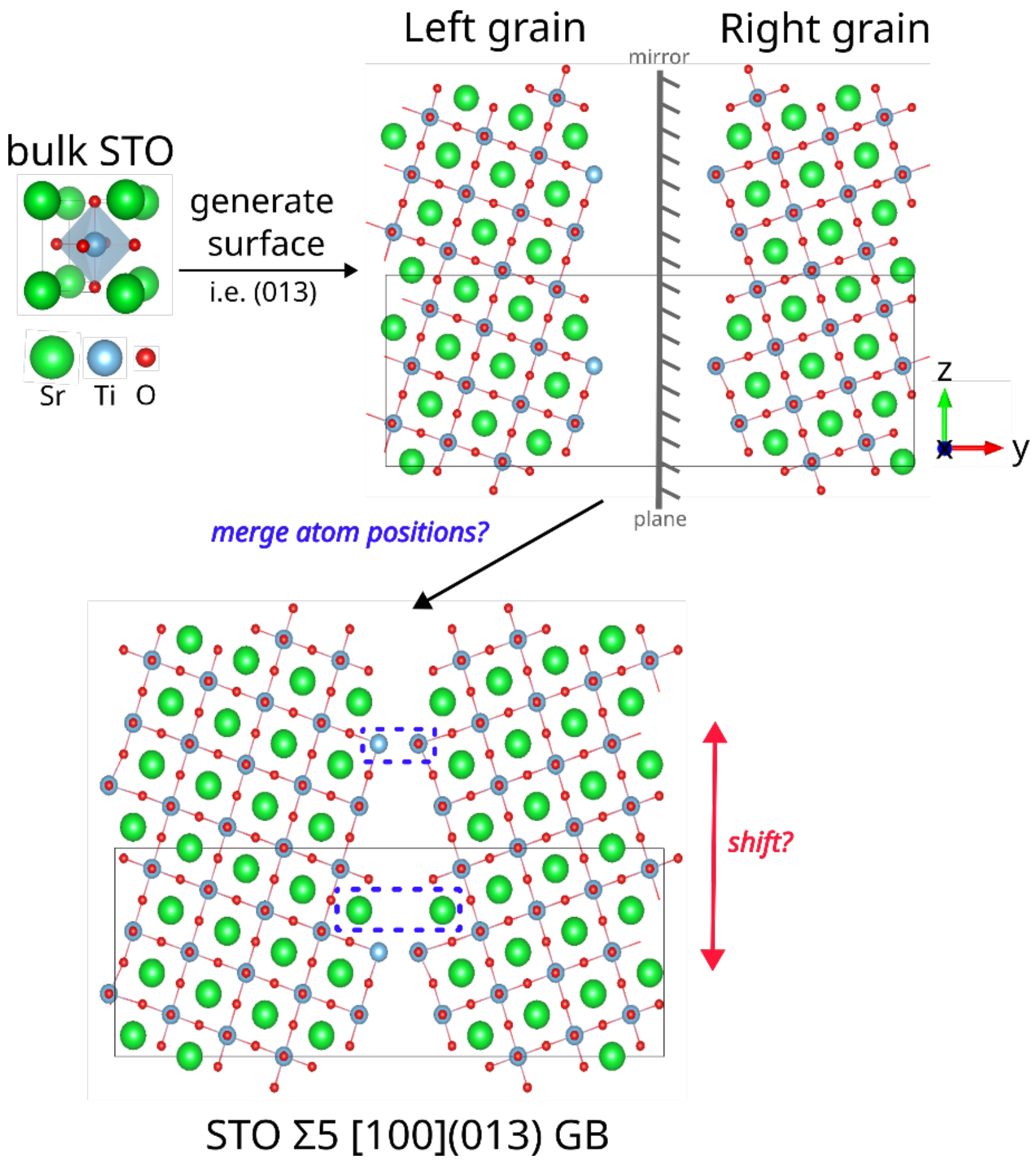

***Figure 2:*** *Schematic of STO Σ5 [100] (013) boundary generation from a bulk unit cell. The grains are mirrored across the yz-plane and the solid rectangle refers to the GB unit cell.*

Using the thus-generated structure, which already showed a Sr-depletion at the boundary as starting point, the optimal composition at the GB was evaluated by successively removing additional Sr atoms from the boundary. For this procedure, a periodically extended system consisting of two GB unit cells and a total of 208 atoms was used, which resulted in an additional Sr-depletion of 2.8 to 11.1% when removing 1 to 4 Sr atoms, respectively. After optimizing the geometry of the different systems, Gibbs free energies were calculated to determine thermodynamic stability according to:

$$\gamma = \frac{1}{2A}\left(E - \sum_i N_i \mu_i\right), \tag{1}$$

where $A$ is the area of the GB or surface, $N_i$ is the number of atoms of type $i$ in the system and $\mu_i$ is the corresponding chemical potential. A description of the calculation approach, including definitions of boundary and surface areas is provided in Section 1 in the supporting information.

## 3. Results

### 3.1. STEM Study of the STO Grain Boundary

HAADF STEM images in Figure 3 reveal a sharp grain boundary composed of symmetric (013) and (012) facets, as well as asymmetric facets with orientations between (071)/(011) and (010)/(043) and lengths from 2 to 10 nm. The symmetric (013)/(013) and the asymmetric facets occur most frequently. The deviation of the asymmetric boundary plane from a nominal symmetric (013) by ~ 9.6° is accommodated by the insertion of (012) asymmetric facets. A qualitative inspection of the HAADF STEM images suggests that the variations in orientation of the asymmetric facets are due to the presence of atomic steps. For details see Figure S5). Occasionally, short symmetric (012)/(012) facets are also found between two (013)/(013) facets. The sharpness of the facets and the absence of Moiré fringes in Figure 3b and c suggest that the facets in this predominately tilt grain boundary are aligned parallel to the tilt axis (*x*-axis), in which case the boundary profile remains unchanged through the lamella thickness (see Figure 1 and inset of Figure 3a).

The relative shifts of the two grains were determined by overlaying the (100) and (110) planes. Shifts in both the y- and z-directions were observed, with magnitudes ranging from 0.01 to 0.1 nm, varying across different positions. A statistical analysis of 17 measurements revealed that the average translation in both the y- and z-directions was 0.04 ± 0.03 nm. The large deviation in these measurements is primarily due to scanning drift during the HR-STEM image acquisition. Literature on symmetric, single-facetted twin boundaries reports shifts less than 0.1 nm in TEM-based studies.[15]

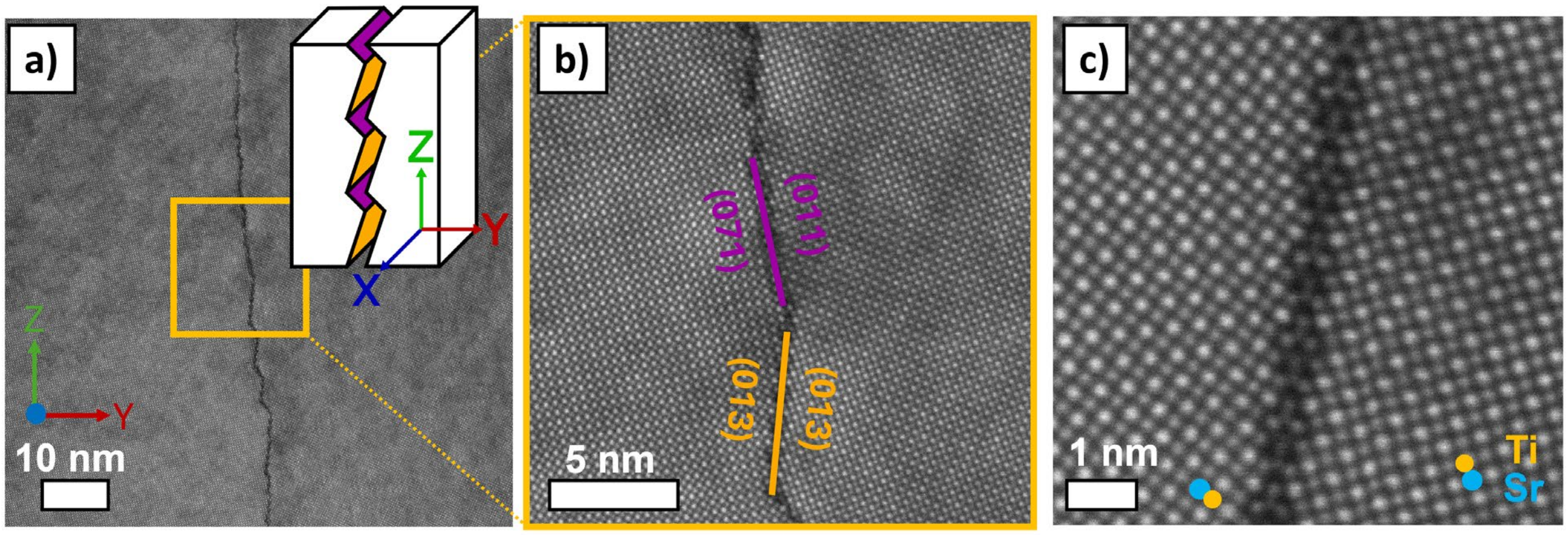

***Figure 3:*** *HAADF STEM of the STO grain boundary.* ***a)*** *STEM image of the Σ5 [100] STO bicrystal in [100] zone axis including an inset showing the proposed facet structure.* ***b)*** *Close-up of a symmetric (013)/(013) facet and an asymmetric facet with close to (071)/(011) orientation.* ***c)*** *High resolution STEM image showing the atomic structure of the symmetric (013)/(013) facet.*

EELS line scans were acquired and used to create composition profiles perpendicular to the boundary. A total of 11 distinct locations along the boundary were successfully investigated by STEM-EELS (see Figure S6 and S7). Representative STEM images and EELS spectral profiles, illustrating the observed behaviors, are shown in Figure 4. All spectra were acquired with a step size less than 0.3 Å. After binning, spatial resolution of approximately 2.5 Å is inferred from the Sr profile. The spectra reveal Sr depletion over a width of approximately 1 nm at the boundary, with depletion levels ranging between 35% and 55%. However, the Ti distribution is heterogeneous. Measurements at five symmetric (013) facets consistently show Ti enrichment of 10% to 40%, with some cases also indicating oxygen enrichment (Figure 4a). In contrast, three measurements at asymmetric facets show no significant change in Ti concentration at the boundary (Figure 4b). Four measurements on the short symmetric (012)/(012) facets show either Ti depletion (Figure 4c) or minor changes in boundary Ti concentration of less than 10%, which is near the signal-to-noise limit of the Ti L-edge signal for these measurements. Furthermore, systematic errors in the determination in the ratio of light and heavy elements such as Ti and O under strong channeling conditions can reach up to 20 percent.[54] Therefore, complementary characterization techniques are necessary to assess the reliability of the EELS data, especially since the channeling conditions vary near the grain boundary and depend on the facets.

The oxygen concentration signal varies from facet to facet and also shows different concentrations in the grains on either side of the boundary (Figure 4c), suggesting the presence of artefacts from multiple scattering and channeling effects. It is well known that inelastic signal modulations on the atomic scale are less pronounced for lighter elements due to stronger channeling, which makes oxygen quantification using EELS particularly

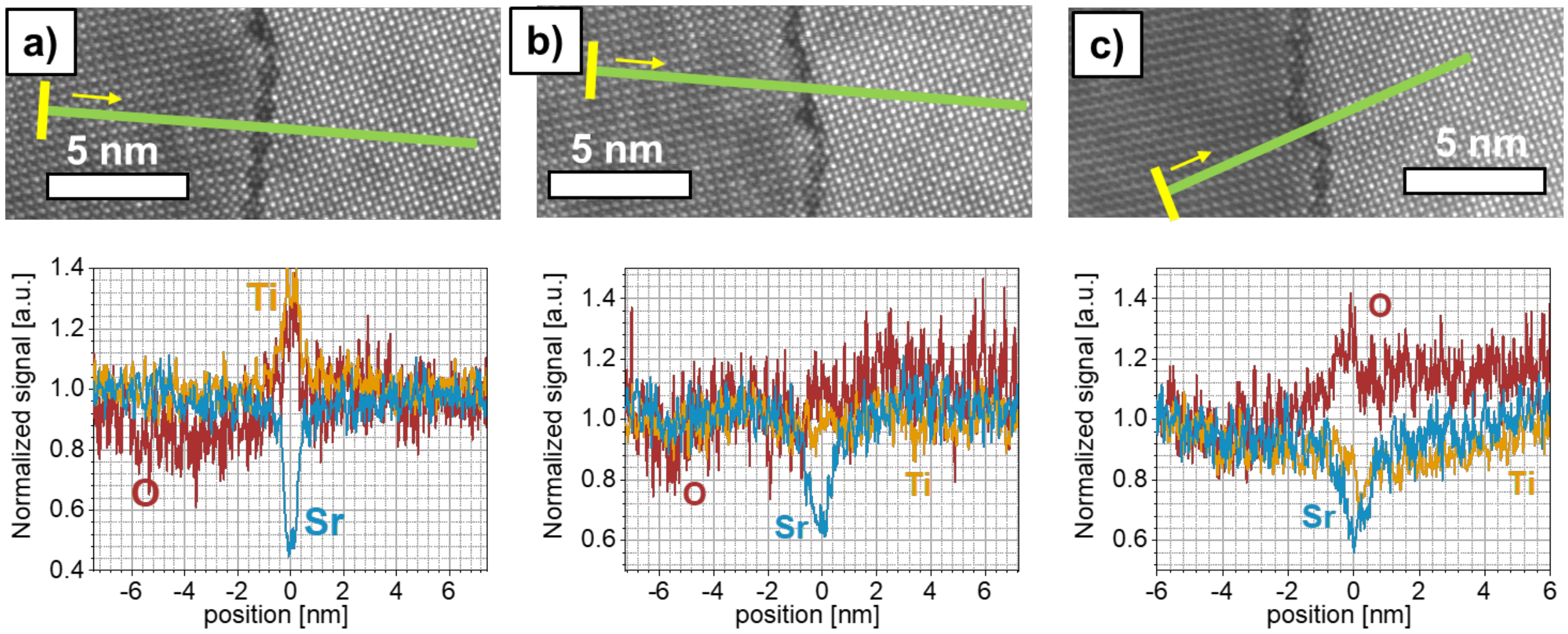

***Figure 4:*** *STEM images (top) and normalized EELS spectra profiles (bottom) from different locations along the STO grain boundary.* ***a)*** *Symmetric (013) facet showing 50% Sr depletion and 30% Ti enrichment.* ***b)*** *Asymmetric (010)/(043) facet showing 35% Sr depletion and no significant change in Ti concentration.* ***c)*** *Short symmetric close to (012)/(012) facet showing 35% Sr and 20% Ti depletion. The yellow bars represent the 2 nm averaging width perpendicular to the profile, while the green lines indicate the length of the EELS profiles*

challenging.[18,21] Therefore, we refrain from interpreting the small changes in measured oxygen signals as evidence of changes in oxygen stoichiometry.

The actual width of the non-stoichiometric boundary layer is likely less than 1.5 nm, as the averaged composition profiles are broadened by the grain boundary roughness and potential grain boundary heterogeneities along the electron beam direction.

3.2. APT Study of STO Grain Boundary

APT specimen fabrication, measurement, and analysis were successfully performed on several specimens containing the STO bicrystal grain boundary. Figure 5a shows a bright field TEM image of a Cr-coated STO specimen, with the grain boundary aligned along the specimen axis. In the reconstructed dataset, the grain boundary was identified by local increases in hydrogen and water from vacuum chamber residual gas, along with a depletion of Sr. This is evident in 2D maps from a 10 nm thick slice through the grain boundary (Figure 5b-d), which show variations in the Ti/Sr, Ti/O, and H/Sr ratios in an edge-on view. The grain boundary is visible as an increase in the Ti/Sr and H/Sr ratios, while the Ti/O ratio remains unchanged, suggesting that the Ti/Sr increase is likely to result from Sr depletion. Additionally, a gradual increase in the H/Sr ratio, and to a lesser extent in the Ti/Sr ratio, is observed deeper into the reconstructed data (Figure 5b and d). This may result from the marked increase in residual gas detection during the measurement,[55,56] which directly impacts the H/Sr ratio since H stems only from the residual gas. The Ti/Sr ratio is also affected due to the unresolved overlap between $^{48}TiO^{2+}$ and $^{16}O_2^+$, since the $^{16}O_2^+$ signal originates partially from the residual gas and thus contributes to an artificial increase in Ti during the later parts of the measurement.

The 2D map of the H/Sr ratio (Figure 5d) reveals that the diffusion-bonded bicrystal grain boundary is rough and appears to have facets with lengths between 2 and 15 nm, consistent with the HAADF STEM images of the same grain boundary (Figure 3 and 4). It is likely that the longer facets in the APT dataset are multiple smaller facets with similar orientation. Furthermore, the 2D maps are taken from a 10 nm thick slice of the dataset and will average over any changes in the grain boundary profile in the view direction. A 1D composition profile perpendicular to the grain boundary and averaged over a width of 100 nm shows a broad depletion in the Sr signal by around 20% at the average position of the grain boundary. No significant variation in the Ti or O signals could be detected.

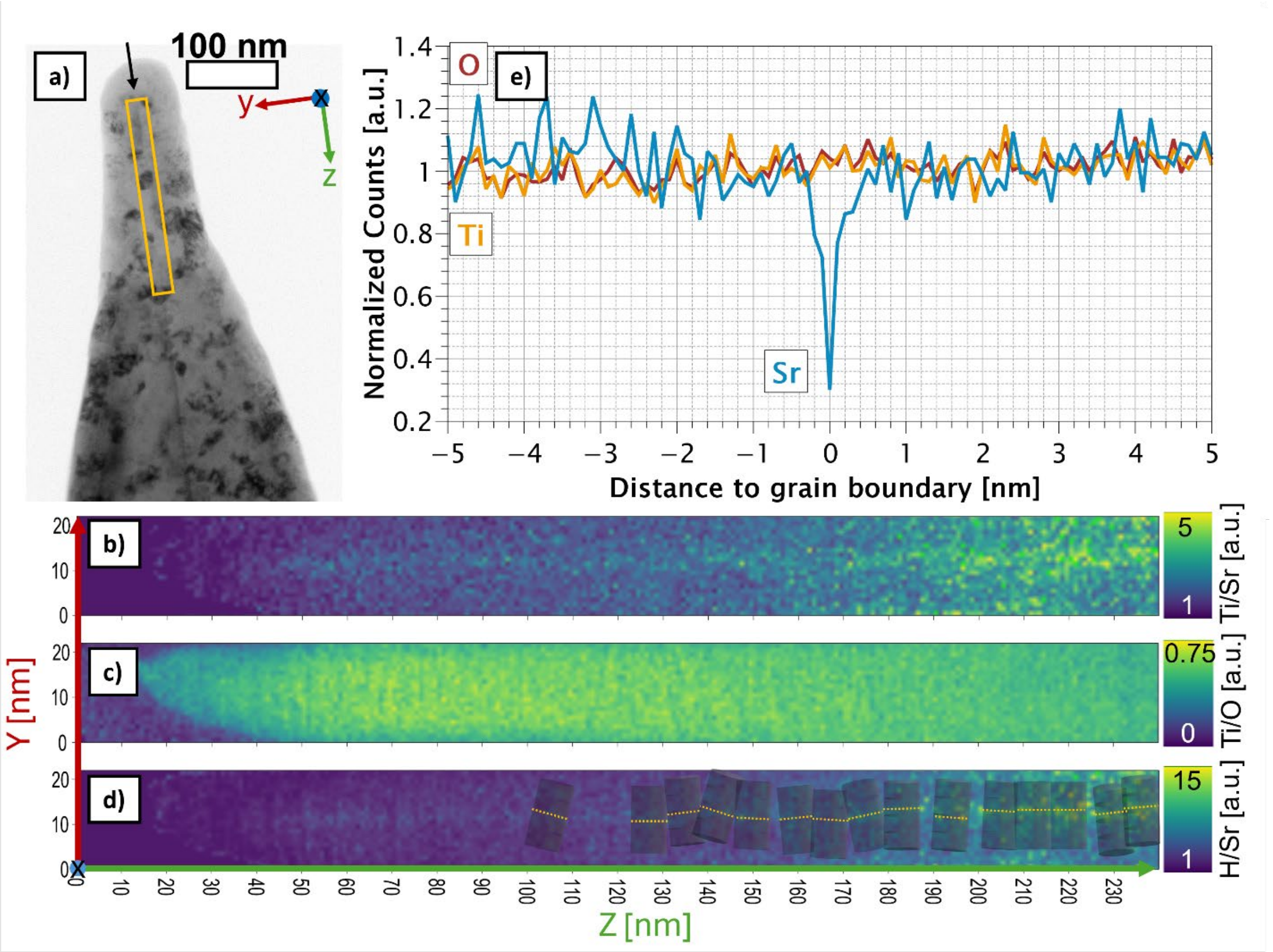

***Figure 5:** APT of the STO grain boundary.* ***a)*** *Bright field TEM image of the Cr-coated STO specimen containing a vertical grain boundary seen in cross-section. The arrow indicates the grain boundary position, and the box indicates the volume shown in the 2D maps in b) – d). The out-of-plane direction is the same as for the lamellae shown in Figures 2 and 3. 2D color maps of the* ***b)*** *Ti/Sr ratio,* ***c)*** *Ti/O ratio, and* ***d)*** *H/Sr ratio obtained from a 10 nm thick slice with 1 nm pixels.* ***e)*** *Averaged Sr, Ti and O counts from individual profiles normalized to bulk values. The profiles were extracted perpendicular to the grain boundary using locally aligned 8 nm diameter cylinders (as shown in d and described in text) and have a resolution of 0.1 nm.*

To account for the faceted nature of the grain boundary, 1D profiles of Sr, Ti, and O were obtained by averaging within 8 nm diameter cylinders, each orientated perpendicular to the respective facets. A clear dip in the Sr signal was detected at a specific orientation for each cylinder, which was used to define the facet orientations to within ±1° (see Section 3 in supporting information). The boundary's approximate location was identified by the coincident positions of H enrichment and Sr depletion. The exact boundary position was then determined by the persistence of the Sr dip over a 1°–2° range of cylinder tilt angles, in contrast to the rapid statistical variations observed in the bulk. The cylinders used for the individual profiles are indicated by shaded regions in Figure 5d, with the boundary locations marked by yellow lines. The normalized averages of counts obtained from the aligned 8 nm cylinder profiles are shown in Figure 5e. These profiles reveal a Sr depletion of approximately 70% in a boundary layer less than 0.5 nm thick, with no detectable Ti or O non-stoichiometry. Some individual cylinder profiles might suggest possible Ti enrichment at some of the facets (Figure S1), but the data scatter is high. Obtaining reliable information about O is particularly challenging, as the Ti and O signals are inherently coupled, evaporating together as various $Ti_xO_y$ molecules (see Figure S2 and Morris et al.[30]). Only a small portion of the detected O is independent of Ti in the $O^{1+}$, $OH^{1+}$ and $O_2^{1+}$ peaks. Of these, the $O_2^{1+}$ is identified as $TiO^{2+}$ for any profiling.

The charge state ratio for $^{48}TiO$ remains unchanged at the grain boundary (see Figure S3), indicating that the Sr depletion and variations in Ti and O are not simply artifacts of local field evaporation effects, such as preferential evaporation, delayed evaporation, or molecular dissociation.[57] Instead, the measured boundary profiles likely reflect true compositional differences at the boundary, indicating clear Sr depletion at the boundary, while neither Ti nor O show significant variations beyond the scatter observed away from the boundary.

The 8 nm cylinders are oriented to be perpendicular to the Sr-depleted region, so they should reflect the local boundary facet normals, provided the facets are at least 8 nm in size. The cylinder axes are broadly distributed around a zero-degree tilt in the *yz*-plane relative to the macroscopic boundary (with the normal along the *y*-axis), with a mean tilt of 5.5° and a maximum tilt of 15°. Tilts in the *xy*-plane are smaller, with 10 out of 15 cylinders having tilts of less than 4°. These findings from the APT analysis suggest that many of the grain boundary facets are close to parallel to the grain boundary tilt axis, as illustrated in the inset of Figure 3a.

### 3.3. DFT Calculations of STO Surfaces and Grain Boundary

Surface energies for (100), (110), (013) and (021) oriented surfaces with stoichiometric and non-stoichiometric compositions have been calculated using the method described in Section 2.4 and S1. The surface energies exhibit significant variation based on facet orientation and atomic composition. The (100):SrO surface termination is the most favorable, exhibiting the lowest surface energy values, while the (100):$TiO_2$ configuration corresponds to the most destabilized surface (Table 1). Table S1 provides a comprehensive list of all the compositions tested with their corresponding surface configurations and calculated surface energies.

| Surface termination | | Atomic Composition | $\gamma_{surf}$ /meV Å$^{-2}$ |
|---|---|---|---|
| (013) | stoichiometric | $SrTiO_3$ | 91 |
| (021) | stoichiometric | $SrTiO_3$ | 100 |
| (100):$TiO_2$ | | $SrTi_{1.2}O_{3.4}$ | 283 |
| (100):SrO | | $Sr_{1.2}TiO_{3.2}$ | -140 |
| (110):$SrTiO_2$ | | $Sr_{1.2}Ti_{1.2}O_{3.2}$ | 231 |
| (110):O | | $SrTi_{1.2}O_{3.4}$ | 341 |

*Table 1: Calculated surface energies $\gamma_{\mathrm{surf}}$ for all investigated surface terminations at 300 K. (The following references were used from the NIST-JANAF database:[58] $\Delta\mu_{\mathrm{Sr}}$ = 0.72 eV, $\Delta\mu_{\mathrm{Ti}}$= 2.32 eV and $\Delta\mu_{\mathrm{O}}$ = −0.27 eV)*

For a given surface facet, Sr depletion and Ti enrichment generally increase the surface energy across all calculated facets. This trend is mainly due to the coordination number of surface atoms. For example, based on the references (*i.e.*, bulk-Ti, bulk-Sr, $O_2$-gas), configurations with fully-coordinated Sr atoms, such as $Sr_{1.3}TiO_{3.2}$, show the lowest surface energy $\gamma_{surf}$ of –146.95 meV·Å$^{-2}$. Overall, compositions with the initially neutral (100):SrO surface termination tend to be highly stable, though Ti enrichment disrupts this trend. The initially stoichiometric and neutrally charged surfaces, (013) and (021), as well as the oxygen-rich (110):O and (110):SrTiO□ orientations, follow similar stability patterns. In contrast, the (100):TiO□ surface turned out to be the least stable one among the studied systems. However, removing Ti atoms or increasing Sr content improves stability, albeit only slightly.

Due to the cuboctahedral coordination of Sr within the crystal lattice, the bulk ligand environment consists of 12 oxygen atoms, with an average bond length $\overline{d}_{\mathrm{Sr\text{-}O}}$ of 2.80 Å, while titanium in bulk STO forms bonds with 6 oxygen atoms at 1.98 Å (Table 2). Not surprisingly, interatomic distances at the surfaces differ slightly from bulk values and depend on the surface composition and termination. The effective coordination numbers at the surfaces show large variations depending on the surface termination and are generally smaller than the bulk values. Due to the step-like layering of the (013) surface, Sr and Ti atoms exhibit two different coordination numbers (*CN*) and different average bond lengths to their adjacent oxygen atoms. However, no differences in Sr–Sr or Ti–Ti distances are observed between the bulk and the analyzed surfaces.

| | | *bulk* | (013) | (021) | (100):$TiO_2$ | (110):SrO | (110):$SrTiO_2$ | (110):O | *Σ5[100](013)* |
|---|---|---|---|---|---|---|---|---|---|
| effective $CN$ | Sr | 12 | 5.8/ 8.6 | 8.6 | 11.6 | 7.8 | 6.7 | 8.9 | 6.6-11.6 |
| | Ti | 6 | 3.9/ 4.9 | 3.9 | 4.9 | 6.0 | 3.9 | 6.0 | 4.7-5.9 |
| $\bar{d}_{\text{Sr-O}}$ / Å | | 2.80 | 2.69/ 2.74 | 2.68 | 2.81 | 2.74 | 2.71 | 2.76 | 2.60-2.79 |
| $\bar{d}_{\text{Ti-O}}$ / Å | | 1.98 | 2.03/ 1.95 | 2.02 | 1.97 | 1.99 | 2.03 | 1.98 | 1.95-2.02 |
| $d_{\text{Sr-Sr}}$ / Å | | 3.97 | 3.97 | 3.97 | 3.97 | 3.97 | 3.97 | 3.97 | 3.85-8.19 |
| $d_{\text{Ti-Ti}}$ / Å | | 3.97 | 3.97 | 3.97 | 3.97 | 3.97 | 3.97 | 3.97 | 3.02-7.10 |

*Table 2: Effective coordination numbers (CN) and average bond length distances for bulk and surface atoms of different terminations and for the Σ5[100](013) grain boundary.*

The structure of the fully-optimized symmetrical Σ5[100](013) grain boundary is shown in Figure 6. The optimized grain boundary structure requires removal of Sr and O atoms due to overlap, resulting in Sr and O depletion at the boundary as illustrated in Figure 7. The averaged concentration profiles of O, Sr and Ti normal to the GB direction reveal Sr depletion of approximately 32% within 0.5 nm and O depletion of roughly 17% within 0.3 nm of the GB. There is no evidence of any Ti depletion or enrichment for this symmetric grain boundary.

The boundary width, determined from the minimum spacing of Sr atoms exhibiting bulk CNs, is approximately 15 Å. A displacement of 1.3 Å in the *z*-direction is observed across the GB, resulting in a mildly shifted kite structure (see Figure 6). This displacement has also been observed in previous TEM studies of symmetrical, single facetted Σ5 boundaries.[15,35,59] The energy of the symmetrical (013) grain boundary was calculated (Equation 1 and Table S3), yielding a value of 149.16 meV·$Å^{-2}$. A comparison with Table 1 shows that this boundary energy falls within the range of some of the surface energies and is somewhat smaller than twice the surface energy of the stoichiometric (013) surface. This supports the experimental observation that diffusion bonding results in relatively low densities of pores in the boundary, as the boundary energy is significantly lower than the surface energies. Additionally, grain boundary energies for further Sr depletion were calculated (Table S3) and found to increase, analogous to the surface behavior (Table 1).

The interatomic distance between strontium atoms across the boundary shows significant variation, ranging from smaller than the bulk spacing (*e.g.*, between Sr1 and Sr4 of 3.85 Å in Figure 6) to much larger values (*e.g.*, between Sr3 and Sr6 of 8.19 Å). Similarly, the innermost Ti atoms in the boundary, T1, T2 and T4, display shorter separations compared to bulk values (Table 2). This decrease in spacing is compensated for by the rearrangement of oxygen atoms, causing additional lattice distortions. Both the *z*-displacement and the lattice distortion reduce the separation between the two grains, thereby enhancing the system stability. Furthermore, the titanium atoms within the grain form bonds with adjacent oxygen atoms, improving the coordination of the titanium atoms.

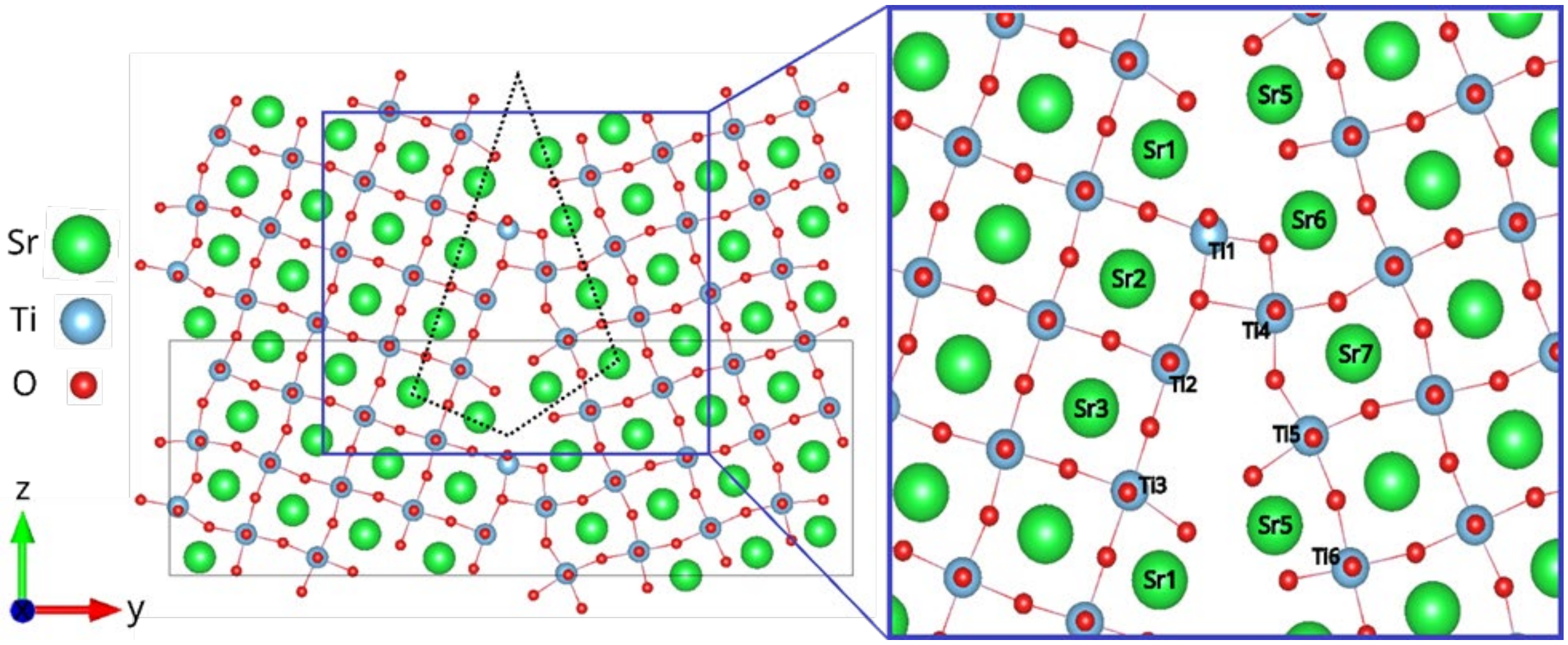

***Figure 6:*** *Stable and fully optimized configuration of Σ5[100](013) strontium titanate GB. The atoms in the close-up (right) in the direct vicinity of the grain are numbered and labelled according to their species, Sr and Ti. The dashed line indicates the distorted kite structure of the grain whereas the solid line showcases the GB unit cell.*

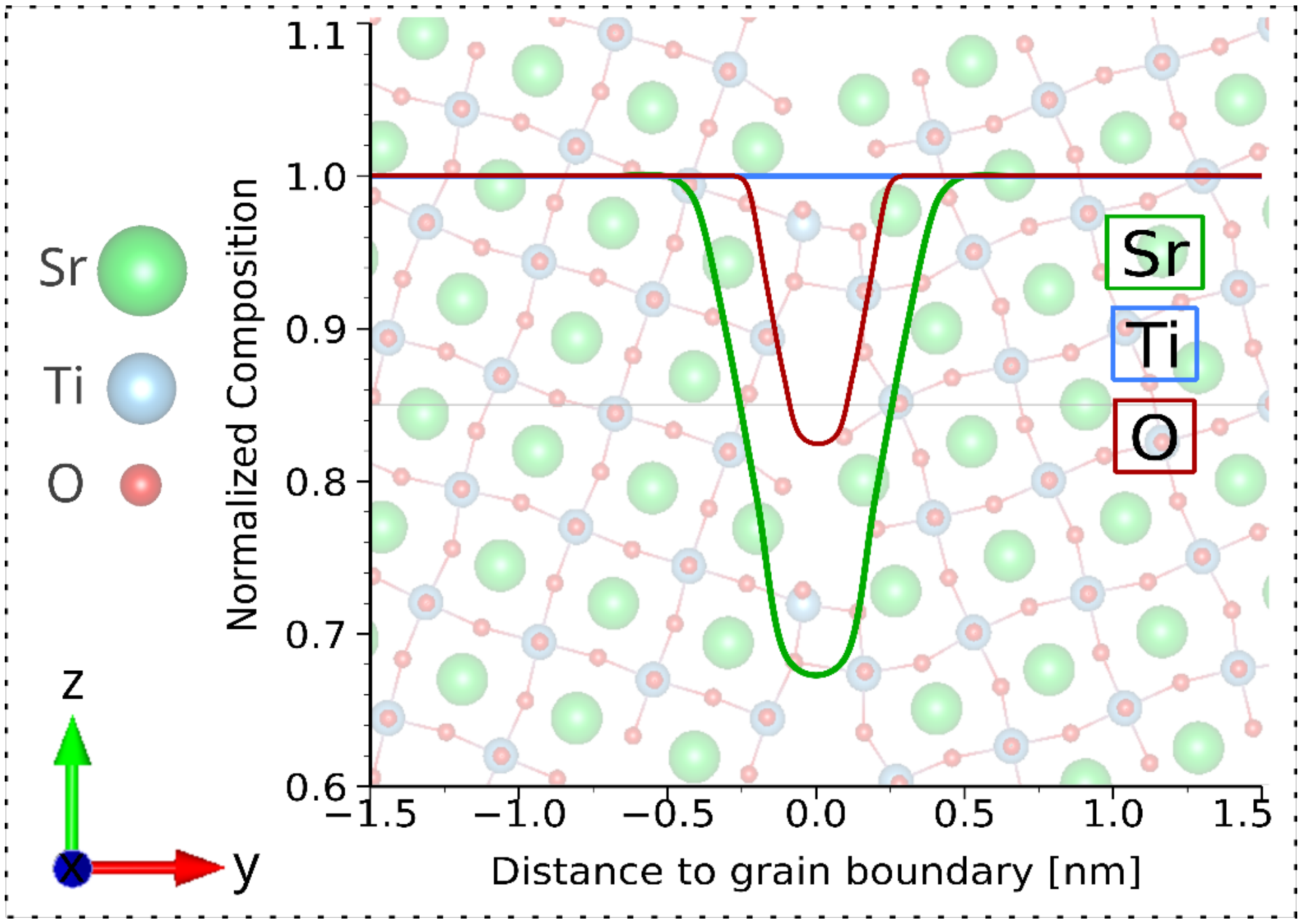

***Figure 7:***. *Averaged Sr, Ti and O concentrations normalized to bulk values for the grain boundary shown in Figure 6. The profiles were extracted perpendicular to the grain boundary by evaluating the corresponding number of atoms in boxes of 0.2 nm width.*

The majority of Sr atoms along the boundary manifest surface-like behavior with regard to average bonding distances and effective *CN*s. Despite the grain comprising two rotated (013) surfaces, only Sr2 and Sr6 (Figure 6) exhibit analogous ligancy and distances to the initial surface. Sr5 displays typical characteristics from the (110):$SrTiO_2$ surface, while Sr4, with a coordination number of 7.3, resembles a Sr atom from the (100):SrO termination, albeit also featuring a reduced average bond distance. The high lattice distortion surrounding the strontium atoms, Sr1 and Sr4, result in a decrease in the average $\overline{d}_{\text{Sr-O}}$ bond lengths to 2.60 Å and 2.64 Å, respectively. Sr3 is the only atom whose effective coordination number approaches its bulk value. In contrast, half of the Ti atoms in proximity to the grain (namely Ti3, Ti5, and Ti6) demonstrate bulk-like properties (Table 2).

## 4. Discussion

Detailed investigations using both APT and STEM of an asymmetric Σ5[100] 36.9° tilt grain boundary reveal that the boundary consists of nanometer-scale symmetric (013), (012) and asymmetric facets.[6] The facet structure, the sharpness of the boundary, and the lack of pores all suggest sufficient mobility during diffusion bonding to allow the structure to reach local energy minima. This is consistent with the observation of both (013) and short segments of (012) facets, which are highly coincident twin boundaries and are expected to have low energies.

STEM-EELS analysis reveals that all investigated boundary facets exhibit Sr depletion of approximately 30% to 50% over a width of less than 1 nm (Figure 4). APT analysis confirms this depletion, and suggests that Sr depletion can be even more pronounced, reaching up to 70% within a boundary width of less than 0.5 nm (Figure 5e). STEM-EELS also shows evidence that Ti is heterogeneously distributed within the grain boundary, with regions of both enrichment and depletion. Notably, Ti is consistently enriched by more than 10% at symmetric (013) facets, which agrees with previous TEM studies that reported Ti enrichment at symmetric Σ5 boundaries.[15] For other facets, STEM-EELS indicates that Ti concentrations are either unchanged or depleted relative to the bulk. However, APT analysis of local facet concentrations does not reveal clear evidence of differences between the bulk and boundary Ti concentrations, suggesting that any differences are likely less than the local resolution of around 10% (Figure S1).

There are artifacts associated with both experimental methods, providing one of the main motivations for complementary characterization techniques. EELS signals depend strongly on channeling conditions, resulting in systematic errors in concentration ratios as large as 20 percent.[54] Similarly, APT measurement accuracy is affected by changes in local evaporation fields at the tip surface due to compositional heterogeneities. Both experimental[60]and modeling[61] studies identify evaporation field artifacts leading to positional inaccuracies on the nanometer scale and compositional inaccuracies of several percent. Although changes in the evaporation field are not detected at the macroscopic level of the grain boundary plane (Figure S3), possible artifacts in boundary width and composition due to variations on the individual facet level cannot be entirely ruled out. Moreover, the disturbed lattice geometry of grain boundaries alone has been shown to introduce significant reconstruction artifacts, even in chemically homogeneous materials evaporating with a uniform evaporation field.[61,62]

Combining the results from the two methods and considering their limitation provides clear evidence for 30-70% Sr depletion over a sub-nanometer width at the grain boundary. Furthermore, Ti heterogeneities of less than 10% deviation from the bulk concentration are likely from facet to facet within the boundary plane. Heterogeneous segregation patterns at grain boundary facets are not uncommon and have been observed in other materials, such as Si[63,64] and stainless steel[65], using a combination of STEM and APT. These patterns have been attributed to the local energy landscape of the boundary, as well as the large strains introduced by the boundary geometry.[64] Future APT and STEM-EELS studies on STO grain boundaries should carefully examine the boundary to further refine the detection of heterogeneities. For example, rotating to boundary so as to lie perpendicular to the APT measurement direction has the potential to improve spatial and compositional resolution .[60–62]

TEM analysis of the images and orientations of the asymmetric facets suggests that they are approximately aligned with (010)/(011) planes. However, their orientations deviate by a few degrees, ranging between (071)/(011) and (010)/(043), due to atomic steps—similar to vicinal surfaces. In contrast, the high quality of the STEM images supports that the symmetric facets are parallel to the electron beam over the electron beam optical depth, which is estimated to be about 20 nm for the selected STEM imaging conditions. This indicates that the symmetric facet normals are perpendicular to the [100] rotation axis used to generate the tilt boundary, resulting in the formation of a low energy facet structure with a simple 2D ledge profile, as illustrated schematically in Figure 1 and the inset of Figure 3a. On the simple assumption that the image quality rules out more than one or two atomic steps along the optical depth at the symmetric facet, the deviation of the symmetric facets from the 2D ledge structure can be estimated as less than 2°.

In apparent contradiction to the TEM results, the APT analysis of facet normals shows that about one-third of the facets deviate by up to 10° from perpendicular to the tilt axis. However, the APT data is quite noisy at the length scale we are trying to resolve here. The resolution perpendicular to the APT analysis direction (which is perpendicular to the grain boundary) is of order 1 nm for homogeneous, defect-free materials. The resolution may be even lower near the grain boundary due to reconstruction artifacts caused by compositional variations and structural disorder at the boundary, as discussed earlier. Therefore, it seems plausible to estimate at least a one nanometer inaccuracy in the positioning and tilt of the 8 nm diameter cylinders used to located the facets, which results in errors of orientation of up to 8°. Thus, we conclude that both methods are consistent with a grain boundary structure dominated by low energy 2D ledge facets (Figure 1 and inset of Figure 3a).

A number of approaches are possible to improve STEM and APT spatial and chemical resolution at grain boundaries. For APT, these improvements involve optimizing the orientation of the boundary relative to the measurement direction ,[60–62] the use of impurity markers, and AI-driven techniques[66,67] to refine APT studies of chemical heterogeneities within boundaries. Advances in TEM, such as multi-slice ptychography, offer the potential for obtaining higher-resolution data on local elemental distributions at grain boundaries.[68] Further studies at asymmetric large angle tilt grain boundaries with different plane orientations would help to shed light into the higher complexity of correlations between GB faceting, atomic scale structure and chemistry of such boundaries. This is of high importance for concluding from idealized bicrystal studies to more realistic microstructures in applications.

The DFT surface studies show an increase in surface stabilization when depleted of Ti atoms or enriched with Sr atoms. The (100):SrO surface termination is the most favorable, while the (100):$TiO_2$ configuration corresponds to the least stabile configuration. Similar results have been observed regarding the (001) and (011) surfaces.[69,70] Both Sr depletion and Ti enrichment generally lead to an increase in surface energy for all calculated surface compositions, which contrasts to the behavior of the GB connecting these individual surfaces.

When optimizing the grain boundary (GB) of the (013) facet, Sr and O atoms undergo depletion due to the merging of similar sites as the separation between the two grains is reduced, resulting in an average Sr depletion of approximately 32% and O depletion of roughly 17% within 0.3 nm of the GB (Figure 7). There is also a displacement of 1.3 Å in the *z*-direction. When the Sr content is further reduced to $Sr_{0.7}TiO_{2.9}$, the z-shift disappears, but an additional translation of 0.4 Å in the *y*-direction occurs. These findings suggest that atomic composition strongly influences the three-dimensional rigid-body translations of the grain boundary.

The shifts observed in our DFT calculations for the Sr depleted case align with the shifts estimated from the STEM studies performed here, within error. They also agree with results of Imaelda *et al.*, who investigated various translation states of the symmetric Σ5[100](013) STO grain boundary. Their most stable GB structure exhibited final translations of 4.2 Å and 0.5 Å in the *z*- and *y*-directions, respectively.[34] Additionally,

EELS and *ab-initio* calculations by Kim *et al.* demonstrated intrinsically nonstoichiometric STO grain boundaries[35] which supports the results from our DFT simulations. A comparison of the grain boundary energy of 149.16 meV·Å$^{-2}$ with the (013) surface energy of 238.85 meV·Å$^{-2}$ for the same composition (Table S3) shows that the grain boundary acts as a strong stabilizer of the reduced Sr content. Reductions in Sr levels induce additional dangling bonds at the surface, which are stabilized at the boundary. Lee *et al.* observed a trend of increasing grain boundary energy with a higher total number of dangling bonds in the symmetric tilt grain boundaries they studied.[37] Based on the calculated energies, the formation of strontium defects along the grain boundary is considered unlikely unless stabilized via additional lattice deformation or compensation mechanisms. Consequently, we hypothesize that the grain boundary formation process is responsible for the Sr depletion observed in EELS and APT experiments.

No change in Ti concentration is observed by DFT when optimizing the symmetrical facet of the (013) surface (Figure 7). Additional Ti boundary enrichment or depletion studies were not performed in our DFT investigations. However, since surface studies show that the presence of additional Ti atoms in various surface configurations destabilizes the surfaces, it is assumed that a similar trend occurs in grain boundaries, albeit to a lesser extent due to the enhanced stabilization provided by the intrinsic dangling bonds in the grain. Our DFT surface studies suggest that Ti depletion along the grain is more probable than Ti enrichment; however, further investigations on asymmetric facets and multiple-facet GBs are required to verify this hypothesis. Additionally, research into defect formation energies at grain boundaries compared to the bulk, has indicated that defects in the vicinity of the GB are more likely to form due to significant tensile strains or an increased number of dangling bonds.[37] Furthermore, an enhanced incorporation of oxygen atoms into the grain would raise the coordination number of the grain atoms, thereby reducing the number of dangling bonds associated with Ti atoms and decreasing the Ti reduction. This would cause the behavior of the Ti and Sr atoms to more closely resemble that of the bulk, further stabilizing the grain boundary.

Although the DFT results do not indicate Ti enrichment at symmetric (013) facets, the variations in Ti content calculated at different surfaces suggest that Ti experiences different energies at various boundary facets, which would cause mobile Ti to move between the facets. If Ti motion out of the boundary were hindered by slow lattice diffusion but allowed to move between facets, we would expect heterogeneous segregation patterns of Ti within the boundaries. This suggestion is supported by the observation of a variability of the asymmetric facet orientations and a sub-facet structure of low index planes and atomic steps which might create a local variability of the chemical potential of the Ti atoms. Current APT results indicate that Ti segregation heterogeneities are less than 10%, but there is still potential for significant effects. This highlights how understanding the energy landscape of the boundary opens possibilities for engineering specific boundary structures and stoichiometries.

The non-stoichiometries and shifts at the grain boundary generate local electric fields that, in turn, influence transport both along and through the boundary. It becomes evident that macroscopic information about a grain boundary is insufficient to determine its local structure or transport properties. Heterogeneities within the boundary play a crucial role in determining the properties and performance of the material. The study presented here demonstrates that experimental methods to map out the structure and composition of grain boundaries, coupled with theoretical methods to determine the energy landscape, are now within reach.

## 5. Summary

STO grain boundaries are important for various technological applications and serve as a model system for studying grain boundaries in other materials. An asymmetric large angle tilt grain boundary in STO was successfully investigated using STEM and APT, revealing the formation of low-energy symmetric and presumably high energy asymmetric facets and a significant Sr depletion over a width of less than 1 nm along the boundary. DFT+*U* studies on the symmetric (013) facets confirmed the Sr depletion and supported the notion that this depletion results from boundary stabilization. STEM-EELS also suggests that the distribution of Ti varies locally from facet to facet, particularly indicating Ti enrichment at the symmetric (013) facets. APT limits this enrichment to less than 10% of the bulk value, while DFT calculations suggest energy differences between the facets, driving Ti redistribution and rearrangement at different boundary facets. The varying local structures in the grain boundary provide distinct trapping or segregation sites for impurities and dopants, indicating that impurity distributions can be engineered within the boundary. On one hand, predicting and controlling the transport properties in STO with grain boundaries will require extremely detailed information about the boundary structure, along with robust modeling of spatially varying interfaces and transport models. On the other hand, this also offers a wide range of possibilities for tailoring properties to influence transport across and along grain boundaries, thereby enabling the design of highly controlled material properties.

**Acknowledgements:** This research is funded by the German Science Foundation (DFG) through the research unit FOR 5065 “Energy Landscapes and Structure in Ion Conducting Solids” (ELSICS), project number 428906592. Access to equipment of the “Collaborative Laboratory and User Facility for Electron Microscopy” (CLUE) at University of Göttingen is gratefully acknowledged.

**Supporting information to**

# Atomic-Scale Investigation of an Asymmetric $SrTiO_3$ Grain Boundary

Janina Malin Rybak[1], Jonas Arlt[1], Qian Ma[1], Carmen Fuchs[2], Baptiste Gault[3,4], Timo Jacob[2], Christian Jooss[1], Tobias Meyer[1], Cynthia A. Volkert[1]

## Section 1: DFT surface and boundary generation and energy calculations

In order to assess the thermodynamic stability of the surface terminations, the Gibbs free energy must be considered. This can be expressed in terms of the total energy of the system *E*, as calculated from the DFT simulations via

$$\gamma = \frac{1}{2A}\left(E - \sum_{i} N_{\mathrm{i}}\mu_{\mathrm{j}}\right), \qquad (1)$$

where $N_{\mathrm{i}}$ is the number of atoms *i* in the system and $\mu_{\mathrm{j}}$ the chemical potential of the species *j*. Since a symmetric slab model is used, the Gibbs free energy is normalized to a surface area of *2A*. With the approximation that the bulk chemical potential of STO $\mu_{\mathrm{STO}}^{\mathrm{bulk}}$ can be written as the total energy $E_{\mathrm{STO}}^{\mathrm{bulk}}$ derived from DFT calculations at thermal equilibrium, a bulk stability condition of

$$\mu_{\mathrm{STO}}^{\mathrm{bulk}} = \mu_{\mathrm{Sr}} + \mu_{\mathrm{Ti}} + 3\mu_{\mathrm{O}} \qquad (2)$$

with

$$\mu_{\mathrm{Sr}} = E_{\mathrm{Sr}}^{\mathrm{bulk}} + \Delta\mu_{\mathrm{Sr}},\ \mu_{\mathrm{Ti}} = E_{\mathrm{Ti}}^{\mathrm{bulk}} + \Delta\mu_{\mathrm{Ti}},\ \text{and}\ \mu_{\mathrm{O}} = \frac{1}{2}E_{\mathrm{O_2}}^{\mathrm{bulk}} + \Delta\mu_{\mathrm{O}}(p,T) \qquad (3)$$

can be developed. The chemical potential of the three species Sr, Ti and O also refer to their bulk or, in the case of oxygen, their gaseous phase energy, as evident in Eq. 3. Utilizing the NIST-JANAF database[58] for $\Delta\mu_{\mathrm{O}}(p, T)$ with T = 0K and p = 1atm, the aforementioned stability condition for the surface morphology can be transformed to

$$\begin{aligned}\Delta\mu_{\mathrm{STO}}^{\mathrm{bulk}} &= \Delta\mu_{\mathrm{Sr}} + \Delta\mu_{\mathrm{Ti}} + 3\Delta\mu_{\mathrm{O}} \\ &= E_{\mathrm{STO}}^{\mathrm{bulk}} - E_{\mathrm{Sr}}^{\mathrm{bulk}} - E_{\mathrm{Ti}}^{\mathrm{bulk}} - E_{\mathrm{O}}^{\mathrm{bulk}}\end{aligned} \qquad (4)$$

and determines the Gibbs free energy of formation for the bulk and gaseous phases, respectively. Adding Eq. 4 to Eq. 1, one can determine the surface area of the different configurations and morphologies of STO surfaces which results in

$$\gamma_{\text{surf}} = \frac{1}{2A}\left[E_{\text{STO}}^{\text{slab}} - N_{\text{Ti}}E_{\text{STO}}^{\text{bulk}} \quad -(N_{\text{Sr}} - N_{\text{Ti}})(\Delta\mu_{\text{Sr}} + E_{\text{Sr}}^{\text{bulk}}) \\ -(N_{\text{O}} - 3N_{\text{Ti}})(\Delta\mu_{\text{O}} + E_{\text{O}}^{\text{bulk}})\right] \tag{5}$$

and considers $\Delta\mu_{\text{O}}$and $\Delta\mu_{\text{Sr}}$ as independent.[70,71] However, it is hypothesized that the species do not form any condensate on the surface, nor do they undergo decomposition. As a result, the chemical potential of each species must be lower than the energy of an atom in the stable phase of the corresponding species, leading to the following conditions:

$$\begin{aligned} \Delta\mu_{\text{STO}}^{\text{bulk}} &\leq \Delta\mu_{\text{Sr}} \leq 0, \\ \Delta\mu_{\text{STO}}^{\text{bulk}} &\leq \Delta\mu_{\text{Ti}} \leq 0, \text{ and} \\ \frac{1}{3}\Delta\mu_{\text{STO}}^{\text{bulk}} &\leq \Delta\mu_{\text{O}} \leq 0. \end{aligned} \tag{6}$$

To characterize the grain boundary properties, the grain boundary energy (GBE) is calculated in a manner similar to the surface energy determination:

$$\gamma_{\text{GB}} = \frac{1}{2S}\left[E_{\text{STO}}^{GB} - N_{\text{Ti}}E_{\text{STO}}^{\text{bulk}} \quad -(N_{\text{Sr}} - N_{\text{Ti}})(\Delta\mu_{\text{Sr}} + E_{\text{Sr}}^{\text{bulk}}) \\ -(N_{\text{O}} - 3N_{\text{Ti}})(\Delta\mu_{\text{O}} + E_{\text{O}}^{\text{bulk}})\right]. \tag{7}$$

In this approach, $E_{\text{STO}}^{GB}$ represents the total grain boundary energy derived from DFT calculations and *S* denotes the cross-sectional area of the grain boundary in the *yz*-plane. The factor of 2 in the denominator accounts for the presence of two grain boundaries in the *y*-direction, arising from periodic boundary conditions.

### Section 2: DFT calculations of STO surfaces and the Σ5[100](013) grain boundary

Surface configurations with different atomic compositions were constructed by deleting 1-2 Sr or Ti atoms or a single SrO molecule from the generated surface structure. The surfaces were optimized, and the surface energy was then calculated according to Eq. 5 using $\Delta\mu_{\text{Sr}}$ = 0.72 eV, $\Delta\mu_{\text{Ti}}$ = 2.32 eV and $\Delta\mu_{\text{O}}$ = −0.27 eV at 300 K. It is evident from Table S1 that surface energy values exhibit significant variation based on atomic composition. Stability increases with higher Sr/Ti ratios as the Sr atomic content increases and the Ti atomic content decreases, resulting in negative surface energy values.

***Table S1:*** *Calculated surface energy values $\gamma_{\mathrm{surf}}$ for all investigated surface terminations and variations at 300 K. ($\Delta\mu_{\mathrm{Sr}}$ = 0.72 eV, $\Delta\mu_{\mathrm{Ti}}$= 2.32 eV and $\Delta\mu_{\mathrm{O}}$ = −0.27 eV.)*

| Surface termination | | Atomic Composition | $\gamma_{\mathrm{surf}}$ /meV Å$^{-2}$ |
|---|---|---|---|
| (013) | stoichiometric | $SrTiO_3$ | 91.04 |
| | Ti enriched | $SrTi_{1.1}O_3$ | 226.44 |
| | | $SrTi_{1.2}O_3$ | 356.28 |
| | Ti depleted | $SrTi_{0.9}O_3$ | 47.21 |
| | | $SrTi_{0.8}O_3$ | -0.37 |
| | Sr enriched | $Sr_{1.1}TiO_3$ | 44.14 |
| | Sr depleted | $Sr_{0.9}TiO_3$ | 157.64 |
| | | $Sr_{0.8}TiO_3$ | 238.85 |
| | SrO depleted | $Sr_{0.9}TiO_{2.9}$ | 168.16 |
| (021) | stoichiometric | $Sr_1Ti_1O_3$ | 100.13 |
| | Ti enriched | $SrTi_{1.1}O_3$ | 180.64 |
| | | $SrTi_{1.2}O_3$ | 267.27 |
| | Ti depleted | $SrTi_{0.9}O_3$ | 59.38 |
| | | $SrTi_{0.8}O_3$ | 20.15 |
| | Sr enriched | $Sr_{1.1}TiO_3$ | 190.55 |
| | | $Sr_{1.2}TiO_3$ | 282.51 |
| | Sr depleted | $Sr_{0.9}TiO_3$ | 141.52 |
| | | $Sr_{0.8}TiO_3$ | 199.23 |
| | SrO depleted | $Sr_{0.9}TiO_{2.9}$ | 142.24 |
| (100):$TiO_2$ | | $SrTi_{1.2}O_{3.4}$ | 283.26 |
| | Ti enriched | $SrTi_{1.3}O_{3.4}$ | 453.35 |
| | Ti depleted | $SrTi_{1.1}O_{3.4}$ | 237.44 |
| | | $SrTiO_{3.4}$ | 207.58 |
| | Sr enriched | $Sr_{1.1}Ti_{1.2}O_{3.4}$ | 237.95 |
| | Sr depleted | $Sr_{0.9}Ti_{1.2}O_{3.4}$ | 405.20 |
| | | $Sr_{0.8}Ti_{1.2}O_{3.4}$ | 544.33 |
| | SrO depleted | $Sr_{0.9}Ti_{1.2}O_{3.3}$ | 419.00 |
| (100):SrO | | $Sr_{1.2}TiO_{3.2}$ | -139.81 |
| | Ti enriched | $Sr_{1.2}Ti_{1.1}O_{3.2}$ | 116.29 |
| | | $Sr_{1.2}Ti_{1.2}O_{3.2}$ | 318.48 |
| | Sr enriched | $Sr_{1.3}TiO_{3.2}$ | -146.95 |

| Surface termination | | Atomic Composition | $\gamma_{surf}$ /meV Å$^{-2}$ |
|---|---|---|---|
| | Sr depleted | $Sr_{1.1}TiO_{3.2}$ | -18.40 |
| | | $SrTiO_{3.2}$ | 79.68 |
| | SrO depleted | $Sr_{1.1}TiO_{3.1}$ | -25.12 |
| (110):$SrTiO_2$ | | $Sr_{1.2}Ti_{1.2}O_{3.2}$ | 230.96 |
| | Ti enriched | $Sr_{1.2}Ti_{1.3}O_{3.2}$ | 395.12 |
| | Ti depleted | $Sr_{1.2}Ti_{1.1}O_{3.2}$ | 99.53 |
| | | $Sr_{1.2}TiO_{3.2}$ | -32.32 |
| | Sr enriched | $Sr_{1.3}Ti_{1.2}O_{3.2}$ | 209.25 |
| | | $Sr_{1.4}Ti_{1.2}O_{3.2}$ | 200.57 |
| | Sr depleted | $Sr_{1.1}Ti_{1.2}O_{3.2}$ | 250.36 |
| | | $SrTi_{1.2}O_{3.2}$ | 272.65 |
| | SrO depleted | $Sr_{1.1}Ti_{1.2}O_{3.1}$ | 301.87 |
| | | $SrTiO_{3.4}$ | 109.45 |
| | Ti enriched | $SrTi_{1.1}O_{3.4}$ | 222.75 |
| (110):O | | $SrTi_{1.2}O_{3.4}$ | 340.79 |
| | Ti depleted | $SrTi_{0.9}O_{3.4}$ | 87.14 |
| | Sr enriched | $Sr_{1.1}TiO_{3.4}$ | 27.60 |
| | | $Sr_{1.2}TiO_{3.4}$ | -33.94 |
| | Sr depleted | $Sr_{0.9}TiO_{3.4}$ | 193.66 |
| | | $Sr_{0.8}$ $TiO_{3.4}$ | 303.12 |
| | SrO depleted | $Sr_{0.9}TiO_{3.3}$ | 182.34 |

The high lattice distortion surrounding the strontium atoms, Sr1 and Sr4 in Figure 5, resulted in a decrease in the average bond lengths to 2.60 Å and 2.64 Å, respectively. Sr3 is the only atom whose effective coordination number approaches its bulk value, while half of the Ti atoms in proximity to the grain (namely Ti3, Ti5, and Ti6) demonstrate bulk-like properties. Most Sr atoms along the grain manifest surface-like behavior with regard to average bonding distances and effective *CN*s (see Table S2). Despite the grain comprising two rotated (013) surfaces, only Sr2 and Sr6 exhibit analogous ligancy and distances to the initial surface. Sr5 displays typical characteristics from the (110):$SrTiO_2$ surface, while Sr4, with a coordination number of 7.3, resembles a Sr atom from the (100):SrO termination, albeit also featuring a reduced average bond distance. Variations in atomic distances or coordination numbers influence the system's polarity and lattice distortion. The presence of dangling bonds is indicated by a lower CN than that found in the bulk; these bonds have a concomitant effect on defect formation energies and GBE.

As observed in surface calculations, where reduced Sr concentrations result in higher surface energies, the GBE also increases with diminishing Sr levels from 159.78 to 209.90 meV Å$^{-2}$ (see Table S3.).

***Table S2:*** *Average bond lengths $\overline{d}_{\text{X-O}}$ between the species X and their nearest neighboring oxygen atoms, along with the corresponding effective CN. The specific interatomic distances across the grain are denoted as $d_{\text{X-X}}$.*

| Species $X$ | $\overline{d}_{\text{X-O}}$ / Å | Effective $CN$ | $d_{\text{X-X}}$ / Å |
|---|---|---|---|
| Sr1 | 2.60 | 8.7 | $d_{\text{Sr1-Sr4}}$ = 3.85 |
| | | | $d_{\text{Sr1-Sr5}}$ = 5.00 |
| Sr2 | 2.74 | 8.9 | $d_{\text{Sr2-Sr5}}$ = 4.71 |
| Sr3 | 2.79 | 11.6 | $d_{\text{Sr3-Sr6}}$ = 8.19 |
| Sr4 | 2.64 | 7.3 | |
| Sr5 | 2.71 | 6.6 | |
| Sr6 | 2.71 | 8.4 | |
| Ti1 | 1.95 | 4.7 | $d_{\text{Ti1-Ti4}}$ = 3.02 |
| Ti2 | 1.96 | 4.9 | $d_{\text{Ti2-Ti4}}$ = 3.52 |
| | | | $d_{\text{Ti2-Ti5}}$ = 4.74 |
| Ti3 | 1.99 | 5.8 | $d_{\text{Ti3-Ti6}}$ = 7.10 |
| Ti4 | 2.02 | 5.2 | |
| Ti5 | 1.99 | 5.9 | |
| Ti6 | 1.98 | 5.9 | |

***Table S3:*** *Calculated grain boundary energies $\gamma_{\text{GB}}$ for the (013) symmetric grain and the Sr depletion variations at 300 K. ($\Delta\mu_{\text{Sr}}$ = 0.72 eV, $\Delta\mu_{\text{Ti}}$= 2.32 eV and $\Delta\mu_{\text{O}}$ = −0.27 eV.)*

| Σ5[100](013) GB | $\gamma_{\text{GB}}$ /meV Å$^{-2}$ |
|---|---|
| | 149.16 |
| - 1 Sr | 159.78 |
| - 2 Sr | 171.10 |
| - 3 Sr | 192.27 |
| - 4 Sr | 209.90 |

## Section 3: Additional APT results

Concentration profiles from three of the individual 8 nm diameter cylinders used to determine the grain boundary facet locations and orientations are shown in Figure S4. Mean values and standard deviations for each cylinder and element are indicated. Each data point shows the average composition in a 0.3 nm thick cylindrical slice. All cylinders show depletion of Sr by amounts between 10 and 80% at the grain boundary position, while there are different variations in the Ti and O signals. However, none of the signals, including the Sr signal, show boundary values that are much outside the scatter observed in the volumes away from the boundary. This indicates that the cylinder slice volume of about 15 nm$^3$ is too small to detect the grain boundary heterogeneities with the measurement protocol and equipment used here. Requiring the boundary signal to be 2 times larger than the background scatter for reliable detection,[33,72] we can estimate that only variations larger than 40% of Sr and around 10% of Ti or O can be detected in volumes as small as the 0.3 nm thick cylinder slices.

The average over all 15 cylinders, reflecting an average over a grain boundary area of around 750 nm$^2$ = (27 nm)$^2$, shows a clear Sr depletion but no noticeable deviation in the Ti and O boundary signals (Figure 5e). Applying the same detection limit criterion as for the individual cylinders, suggests that the average Ti and O compositions in the boundary deviate by less than around 10% from the bulk values.

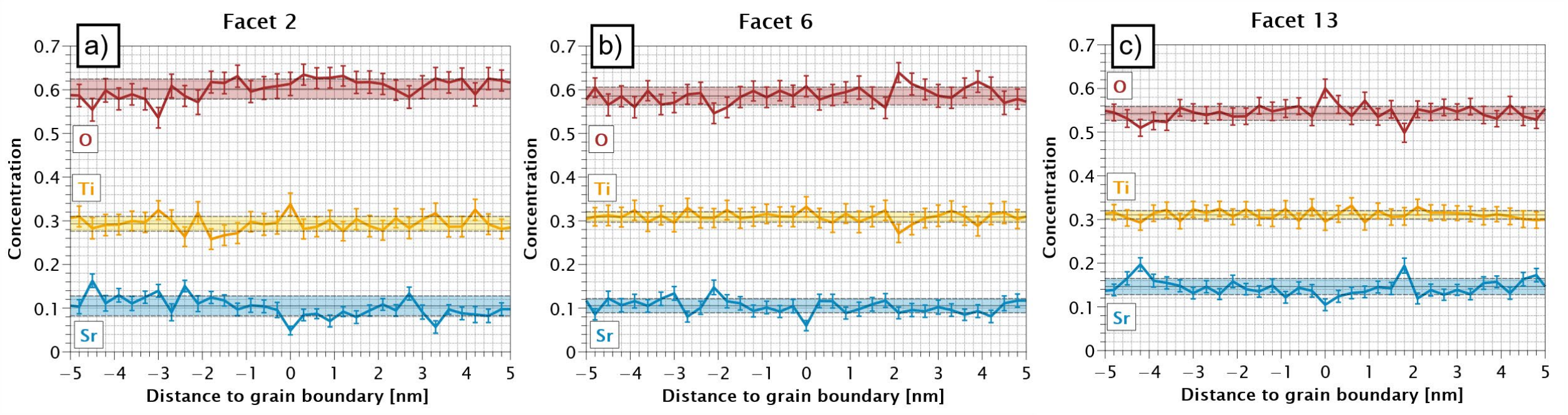

***Figure S1***: *Concentration profiles at selected facets showing* ***a)*** *possible Ti enrichment,* ***b)*** *stoichiometric behavior for O and Ti and* ***c)*** *a possible O enrichment. The concentration profiles have a bin width of 0.3 nm. The shaded regions represent the mean bulk concentrations and standard deviations calculated from 1 to 5 nm away from the grain boundary. The error bars of the profiles represent the standard deviation calculated as described in Danoix et al.* [73,74]

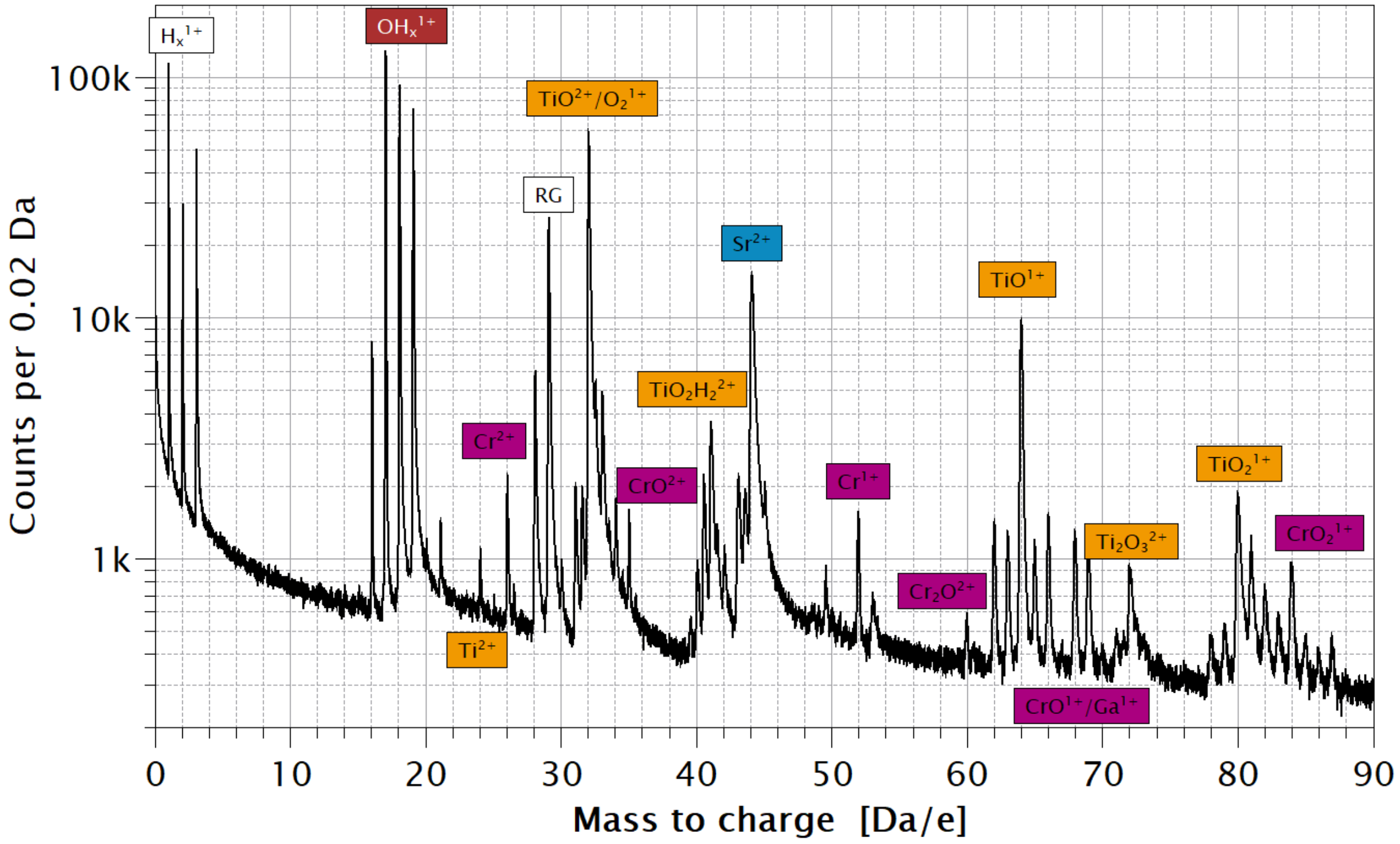

***Figure S2:*** *Mass spectrum of the STO core with most of the Cr coating removed via compositional filtering. Almost all Ti comes from various* $Ti_xO_y$ *peaks. Some O in the form of* $O^{1+}$ *and* $OH^{1+}$ *is independent of Ti.* $OH_s^{1+}$ *and* $OH_3^{1+}$ *originate from the residual gas in the system and are thus excluded from compositional analysis.*

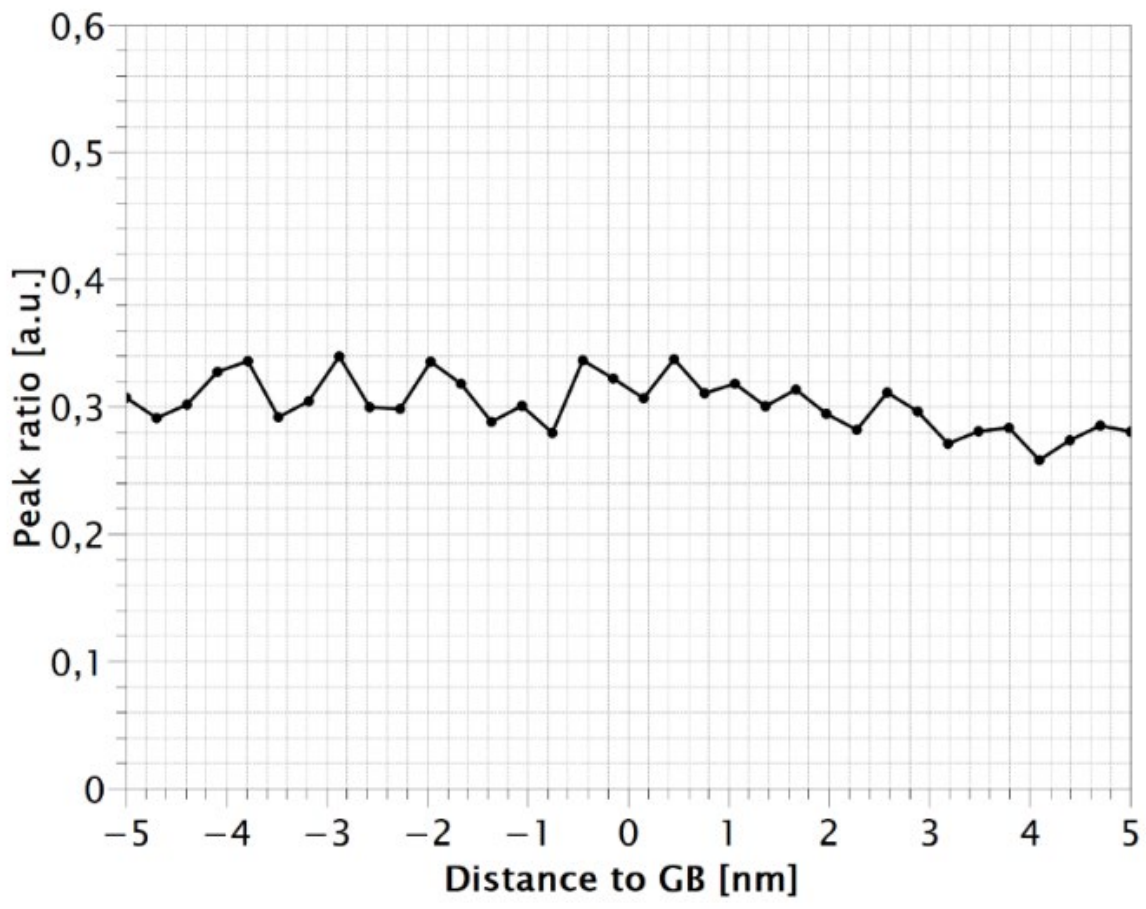

***Figure S3:*** *Ratio of* $^{48}TiO^{1+}$ *to* $^{48}TiO^{2+}$ *across the grain boundary, extracted from a broad profile between z = 100 nm to 200 nm (see Figure 4b-d). The charge state ratio* $^{48}TiO^{+}/^{48}TiO^{2+}$ *remains unchanged at the boundary, indicating that, on the macroscopic scale, local electric fields and evaporation conditions do not vary significantly at the boundary during APT measurements.*

The formation of molecular ions with different charge states is highly dependent on the electric field so that ratios of charge states can be used to detect variations in the electric field at the surface of the APT specimen during analysis.[75,76] Variations in electric field occur if the specimen apex deviates from a spherical shape (assuming that the specimen surface is uniformly conducting) and will develop during measurement if the evaporation fields are inhomogeneous, as can be expected for specimens that are compositionally inhomogeneous. The fact that there is no measurable change at the grain boundary in the ratio of $^{48}TiO^{1+}$ to $^{48}TiO^{2+}$ (Figure S3) indicates that differences in evaporation field at the macroscopic level of the boundary plane remain below the sensitivity of our measurements. However, local variations at individual facets cannot be ruled out.

**Section 4: Additional TEM results**

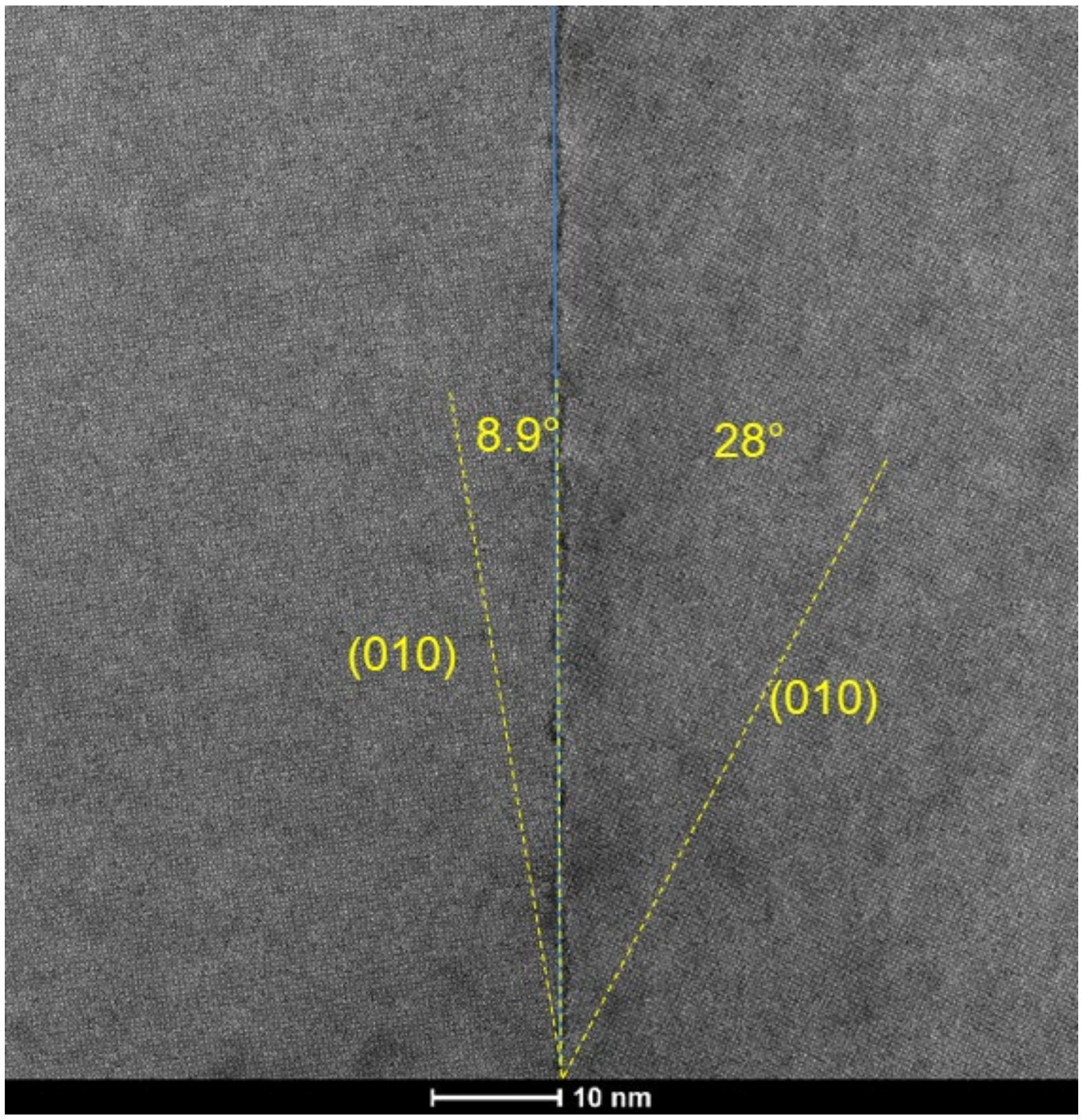

***Figure S4**: HAADF STEM image showing the geometry of the asymmetric grain boundary plane.*

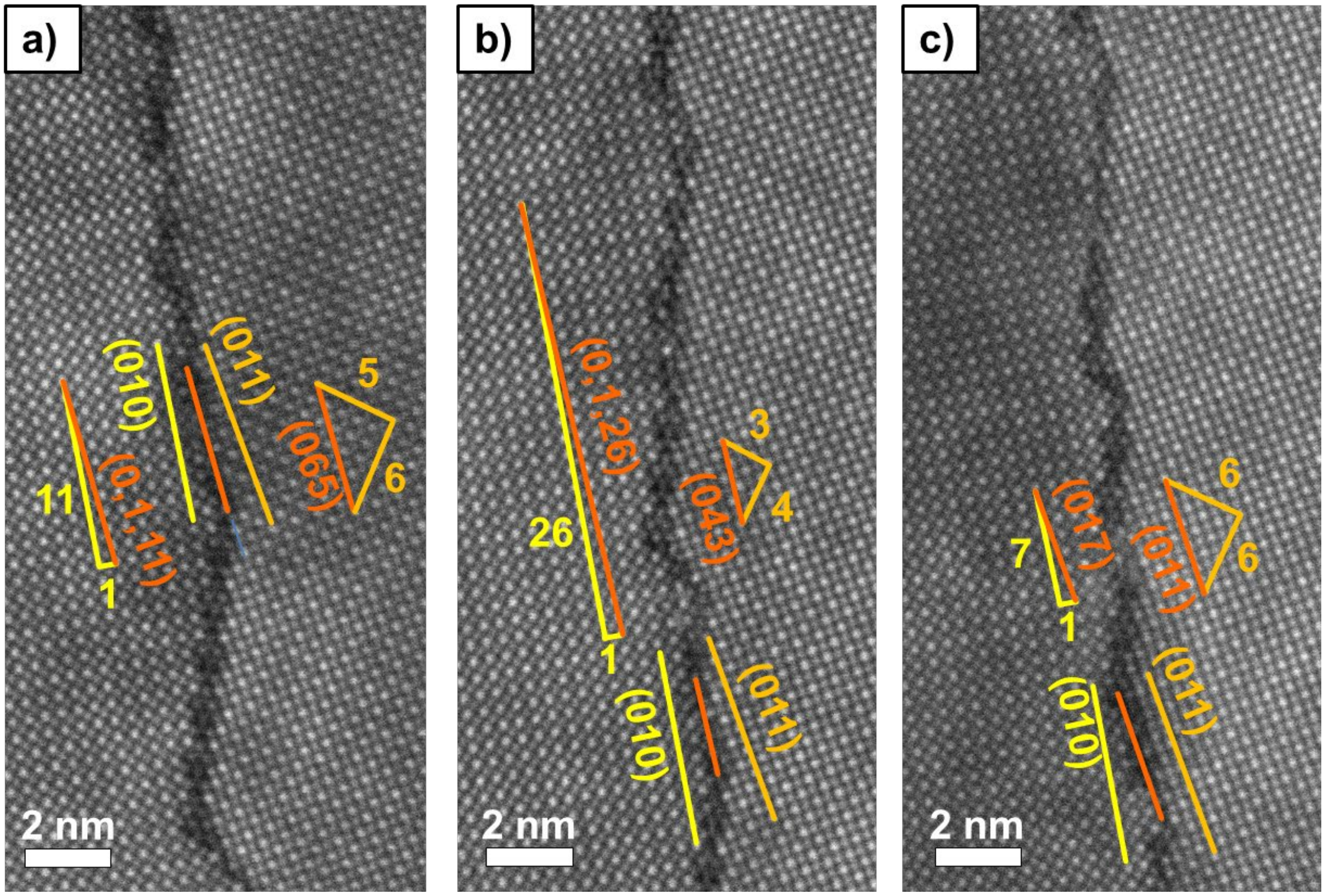

***Figure S5:*** *HAADF HRSTEM images of different GB positions showing a slight variability of the GB plane orientation of asymmetric facets and the corresponding termination planes of both grains.*

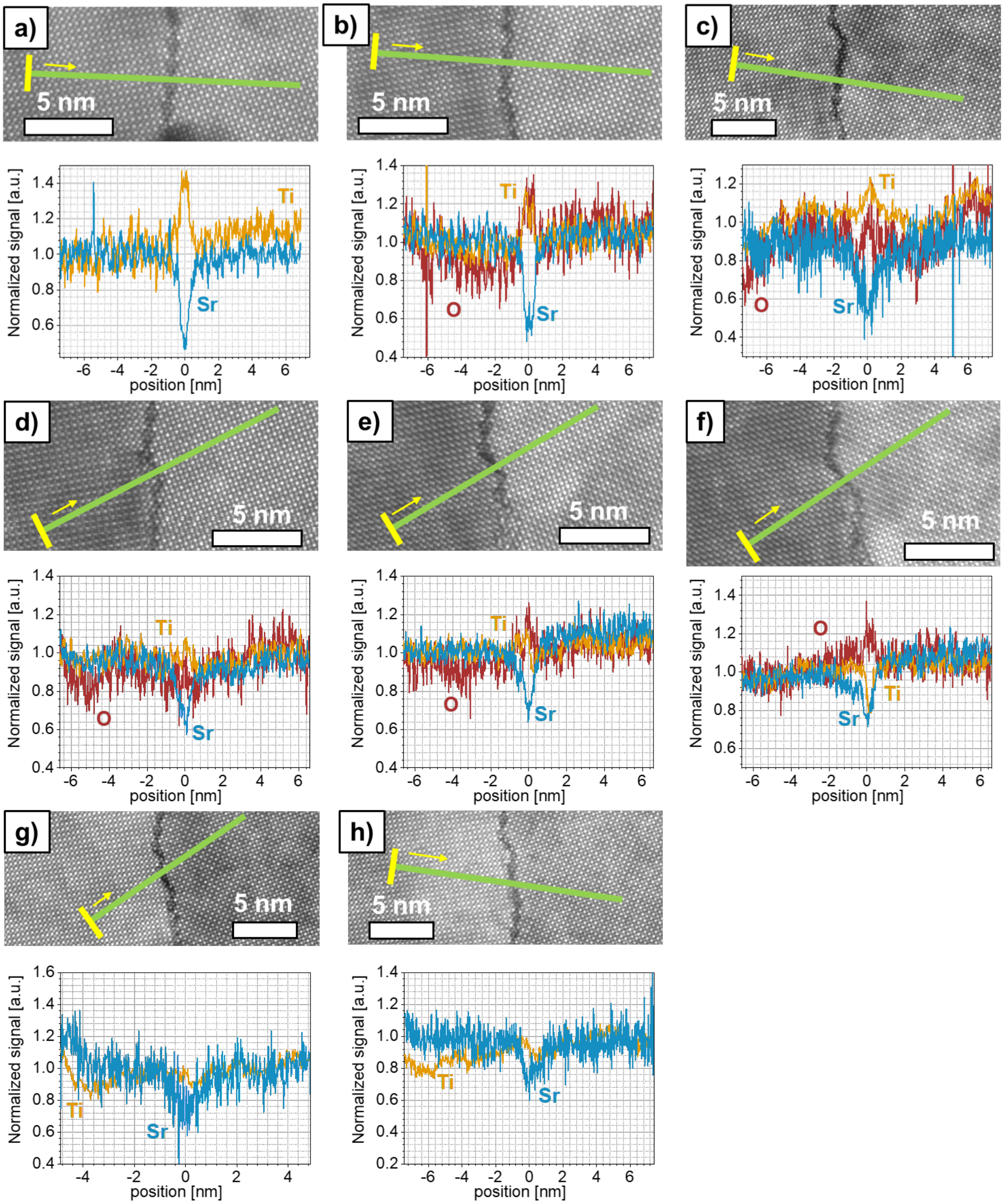

***Figure S6:*** *STEM images and normalized EELS spectra profiles from different locations along the STO grain boundary. a-c) Symmetric (013) facets. d-f) Short symmetric close to (021)/(021) facets. g-h) Asymmetric close to (071)/(043) facets. The yellow bars and green lines show the perpendicular averaging width and length of EELS profiles. In a), g) and h) the oxygen profiles are not shown because of the too high noise level.*

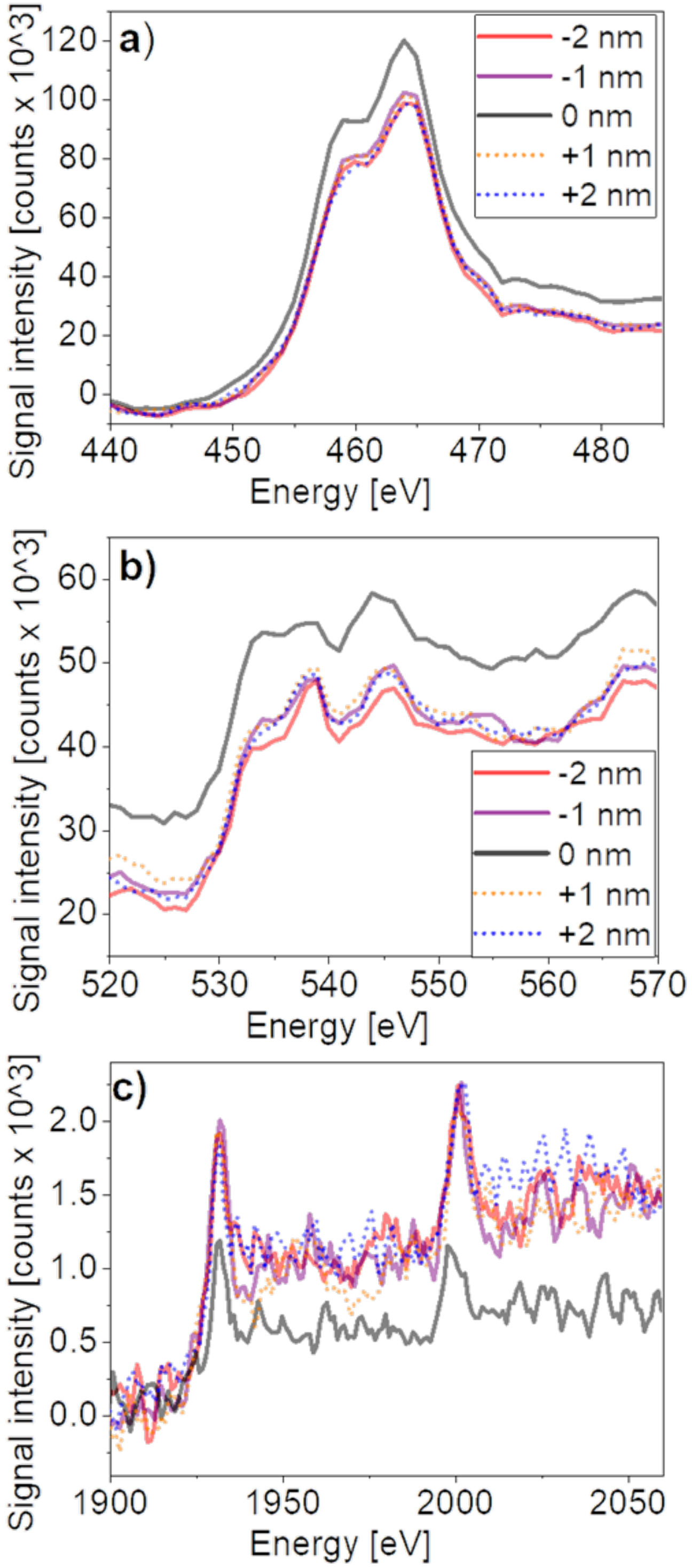

***Figure S7:*** *EELS spectra after background subtraction of a) Ti-L edge, b) O-K edge and c) Sr-L edge from the line profiles shown in Figure 4a). The legend refers to the distance from the GB.*